\documentclass[twocolumn,showpacs,amsmath,amssymb,showkeys,floatfix,prc,aps]{
revtex4}
\usepackage{amsfonts,amsmath}
\usepackage{graphicx}
\usepackage{dcolumn}
\usepackage{bm}

\begin{document}

\title{Implications of the Super-K atmospheric data for the 
mixing angles $\theta_{13}$ and $\theta_{23}$}

\author{J. E. Roa$^1$, D. C. Latimer$^2$,   and D. J. ErnsJt$^1$}

\affiliation{$^1$Department of Physics and Astronomy, Vanderbilt University,
Nashville, Tennessee 37235}

\affiliation{$^2$Department of Physics and Astronomy, 
University of Kentucky, Lexington, Kentucky 40506}

\date{\today}

\begin{abstract}
A three-neutrino analysis of oscillation data is performed using the recent,
more finely binned Super-K oscillation data, together with the CHOOZ, K2K, and
MINOS data. The solar parameters, $\Delta_{21}$ and $\theta_{12}$, are fixed
from a recent analysis, and $\Delta_{32}$, $\theta_{13}$, and $\theta_{23}$ are
varied. We utilize the full three-neutrino oscillation probability and an exact
treatment of the Earth's MSW effect with a castle-wall density.  By including
terms linear in $\theta_{13}$ and $\varepsilon := \theta_{23} - \pi/4$, we find
asymmetric errors for these parameters $\theta_{13}=-0.07^{+0.18}_{-0.11}$ and
$\varepsilon=0.03^{+0.09}_{-0.15}$.  For $\theta_{13}$, we see that the lower
bound is primarily set by the CHOOZ experiment while the upper bound is
determined by the low energy $e$-like events in the Super-K atmospheric data.  We
find that the parameters $\theta_{13}$ and $\varepsilon$ are correlated --- the
preferred 
negative value of $\theta_{13}$ permits the preferred value of
$\theta_{23}$ to be in the second octant, and the true value of $\theta_{13}$ 
affects the allowed region for $\theta_{23}$.
\end{abstract}



\maketitle

\section{INTRODUCTION}

The experimental observation of neutrino oscillations implies that at least two
of three neutrinos have mass, and the mass eigenstates differ from the flavor
eigenstates. The vast majority of oscillation experiments
\cite{sksolar1,sksolar2,homestake,gallex,sage,gno,sno04,sno05,sno07,sno08,
araki:2005,abe:2008,skatm1,skatm2,skatm3,skatm4,k2k05,k2k06,
Adamson:2007,Adamson:2008},
including the null result from the CHOOZ reactor experiment \cite{chooz}, can be
globally understood in terms of three mixing angles $\theta_{jk}$, with
$j=1,2,3$ and $j<k$, one phase $\delta$, and two independent mass-squared
differences $\Delta_{kj}:=m^2_k-m^2_j$. 
The separation between two of the mass-squared differences is sufficiently large
so that the data from a given experiment, which may span some range of baselines
and neutrino energies, can be approximately understood within the context of an
effective two-flavor theory.  
Experiments detecting solar neutrinos
\cite{sksolar1,sksolar2,homestake,gallex,sage,gno,sno04,sno05,sno07,sno08} and
the long baseline (LBL) reactor experiment KamLAND \cite{araki:2005,abe:2008} 
are particularly sensitive to the mixing angle $\theta_{12}$ and the
mass-squared difference $\Delta_{21}$ assuming the standard
representation of the neutrino mixing matrix \cite{Amsler:2008}.
A three neutrino analysis \cite{schwetz:2008} gives a value for the mixing angle
 $\sin^2\theta_{12}=0.304^{+0.046}_{-0.034}$ (2 $\sigma$ error), with a 
precision of 8\% at 3$\sigma$. The solar mass-squared difference is determined
predominantly by the SNO data \cite{sno08} and is found to be $\Delta_{21} =
7.65^{+0.47}_{-0.40}\times 10^{-5}$ eV$^2$.
Atmospheric and accelerator beam stop neutrinos provide experimentalists with a
good source with which to measure $\theta_{23}$ and $\Delta_{32}$.
MINOS \cite{Adamson:2007,Adamson:2008} predominantly determines $\Delta_{32}$
while the mixing angle $\theta_{23}$ is determined mainly by the Super-K
atmospheric data \cite{skatm1,skatm2,skatm3,skatm4}. Present values for these
parameters \cite{schwetz:2008} are $\Delta_{32}=2.40^{+0.24}_{-0.22}\times 10^{-3}$ 
eV$^2$ and $\sin^2\theta_{23}=0.50^{+0.13}_{-0.11}$.
The remaining mixing angle, $\theta_{13}$, mixes the two scales.  This same
analysis gives $\sin^2\theta_{13}\leq 0.040$; recent analyses hint at a value of
$\theta_{13}$ differing from zero
\cite{Fogli:2008,Balantekin:2008,Maltoni:2008}. Recent
review articles can be found at Refs.~\cite{Gonzalez:2008,Fogli:2006}.

As we enter the era of precision measurements, global analyzes of neutrino data
must employ a full three-neutrino framework in order to correctly assess the
neutrino mixing parameters.  This will become evident herein as we consider
various experiments' impact upon the small parameters $\theta_{13}$ and
$\varepsilon := \theta_{23} - \pi/4$, the deviation of $\theta_{23}$ from
maximal mixing. The quantitative knowledge of $\theta_{13}$ is a particularly
important part of neutrino oscillation phenomenology because it sets the
magnitude of possible CP violating effects as well as the size of effects that
might be used to determine the neutrino mass hierarchy. There are presently
three new reactor experiments planned or under construction which are designed
to measure $\theta_{13}$, Daya Bay \cite{Guo:2007}, Double CHOOZ \cite{dchooz},
and RENO \cite{reno};
an LBL experiment is also planned, T2K \cite{t2k}. The subsequent generation of
experiments, which will be designed to ascertain the level of CP violation,
cannot proceed until the current generation better determines the value of
$\theta_{13}$. In addition, a more quantitative knowledge of the mixing angles,
and particularly of $\theta_{13}$, can help discern between models and
symmetries of the physics that underlies neutrino mixing \cite{Raidal:2008}. 
The deviation of $\theta_{23}$ from maximal mixing is also important in model
building as it might indicate the presence of a broken symmetry.  At short
baselines, the oscillation probabilities which might probe the mixing angle
$\theta_{13}$ are quadratic in this small parameter; however, we 
have previously shown there are terms in the oscillation probability
which are linear and appreciable at very long baselines (VLBL)
\cite{Latimer:2005a,Latimer:2005b,Latimer:2005c} and arise from interference 
between the oscillations driven by
the two mass-squared differences.  This is also a region of the parameter space
where one can look \cite{Latimer:2007,Peres:2004} for CP violating effects.
The sub-GeV data set of the Super-K atmospheric experiments is 
potentially sensitive to such effects.
Furthermore, it was shown in Ref.~\cite{Latimer:2005c} that there is a
non-trivial relation between $\varepsilon$ and $\theta_{13}$ for sub-GeV
neutrinos at VLBLs.  As such, the extraction of these parameters from the
atmospheric data requires a full three neutrino treatment since approximations
overly simplify the correlations of the parameters.

We here investigate atmospheric neutrino oscillations with the full three
neutrino oscillation probabilities. Because we do not use truncated expansions, 
all terms linear in $\theta_{13}$ and $\varepsilon$ will be considered as well
as higher order contributions.  We do not expect a large change in the extracted
parameters as only a limited number of the Super-K data bins lie in the region
where linear terms will be significant. On the other hand, in the context of
atmospheric data, $\theta_{13}$ is itself a small effect as is the octant of
$\theta_{23}$. Small effects can sometimes have a proportionally larger impact
on something that is inherently small. In keeping with the use of the full three
neutrino oscillation probabilities, we also utilize the method proposed in
Refs.~\cite{Ohlsson:1999,Ohlsson:2000} to treat the MSW effect
\cite{msw1,msw2}. 
With a
castle wall density profile of the earth, this treatment of the MSW effect is
exact so that approximate expressions for the oscillation probabilities are not
needed. We also include a model for the multi-ring events, a data subset often
neglected.

\section{Analysis}

In vacuo, the probability that a neutrino of flavor $\alpha$ and energy $E_\nu$
will be detected as a neutrino of flavor $\beta$ after traveling a distance $L$
is given by
\begin{equation}
{\mathcal P}_{\alpha \beta}(L/E_\nu) = \delta_{\alpha \beta} - 4 
\sum^3_{\genfrac{}{}{0pt}{}{k <j,}{j,k=1}} (U_{\alpha j} U_{\alpha k} U_{\beta
k} 
U_{\beta j}) \sin^2 {\varphi_{jk}}\,\,,
\label{exact}
\end{equation}
with $\varphi_{jk} := 1.27\,\Delta_{jk}\,L/E_\nu$, where $L$ is measured in
kilometers, $E_\nu$ in GeV,
and the mass 
eigenvalues $m_i$ in eV. 
The matrix $U_{\alpha i}$ is the unitary matrix that relates the mass basis $i$
to the flavor basis $\alpha$. We assume CP conservation so that the $U_{\alpha
i}$
are real.   Neutrinos which propagate long distances through matter of
sufficient densities can incur significant interactions which are diagonal in
flavor.  For matter of constant density, the upshot of these interactions is a
modification of the effective mixing angles and mass-squared differences so that
an oscillation formula similar to Eq.~(\ref{exact}) holds.  The density of the
earth may be approximated as piecewise constant \cite{prem}. In addition, for
certain energies and densities, the neutrinos can undergo parametric resonances
in regions of varying densities \cite{Akhmedov:1998}.  To account for these
interactions, we employ a simple model of the earth:  a mantle of density 4.5
gm/cm$^3$ and a core of density 11.5 gm/cm$^3$ with  radius 3486 km.  Using the
methods in Refs. \cite{Ohlsson:1999,Ohlsson:2000}, we are able to fully
incorporate an exact three neutrino model of the neutrino-matter interactions
which automatically incorporates any possible parametric resonances.

Our interest is to study and extract the following parameters from the
experimental data: $\theta_{13}$, $\theta_{23}$, and $\Delta _{32}$. As such, we
fix the solar mixing parameters from a recent analysis \cite{Gonzalez:2008},
$\theta_{12}=0.58$ and $\Delta_{21}=8.0\times 10^{-5}$ eV$^2$. Given that there
is no evidence to indicate CP violation, we assume CP conservation and take
$\delta=0$. We include the details of our analyses of the relevant experiments
in the Appendix.  We comment on them briefly here.
The Super-K atmospheric data is statistically the most significant data set, and
it covers a range of over four orders of magnitude in $L/E$.  Our analysis
employs the most recent more finely binned data  \cite{skatm4}, a necessity for
studying the small parameters $\theta_{13}$ and $\varepsilon$.  We also include
the most recent MINOS results \cite{Adamson:2008}, the K2K results \cite{k2k06},
and the CHOOZ results \cite{chooz}.

In order to ascertain the importance of the linear and higher-order terms in
$\theta_{13}$ (and also $\varepsilon$), we compare our results with those
generated by the often used sub-dominant approximation which arises from an
expansion in the ratio of the mass-squared differences,
$\alpha\equiv\Delta_{12}/\Delta_{32}$. In this approximation, the leading order
oscillation probabilities are given by
\begin{eqnarray}
{\mathcal P}_{ee}&=&1-\sin^22\theta_{13}\,\sin^2\left(\varphi_{32}\right)
\nonumber \\
{\mathcal P}_{e\mu}&=&
\sin^2\theta_{23}\,\sin^22\theta_{13}\,\sin^2\left(\varphi_{32}
\right) \nonumber \\
{\mathcal
P}_{\mu\mu}&=&1-4\cos^2\theta_{13}\,\sin^2\theta_{23}\,(1-\cos^2\theta_{13}\,
\sin^2\theta_{23})
\nonumber \\
&&\times \sin^2\left(\varphi_{32}\right)\,\,. 
\label{subd}
\end{eqnarray}
Additional correction terms \cite{expans,Gonzalez:2008,Fogli:2006}
can then be added. The results for the sub-dominant approximation when compared
to the results for the full three-neutrino oscillation probabilities will inform
us of the size of the correction terms.

\begin{figure}
\includegraphics[width=3.3in]{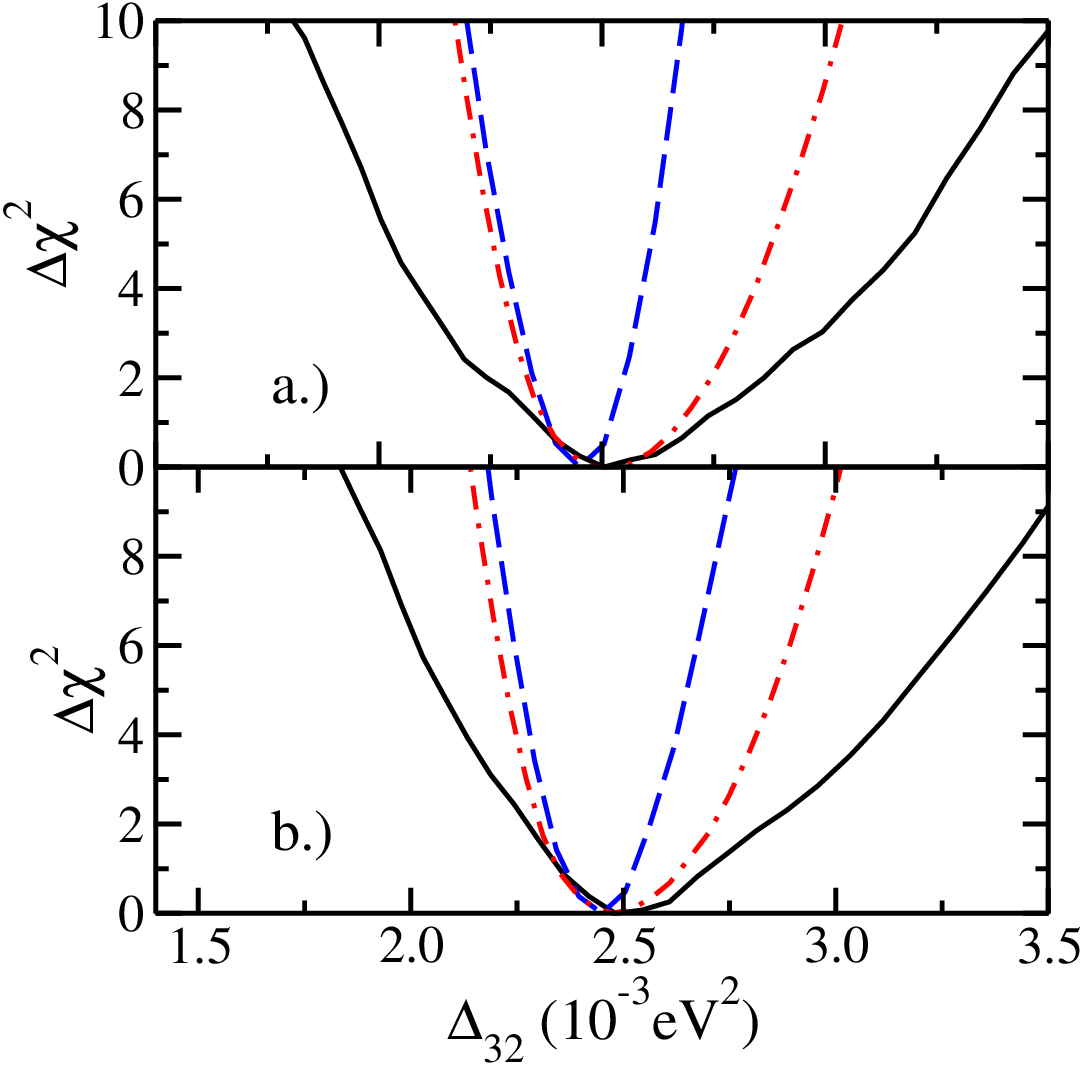}
\caption{[color online] $\Delta\chi^2$ versus mass-squared difference
$\Delta_{32}$ for the a.) sub-dominant approximation and b.) full three-neutrino
calculation. The [black] solid curves utilize only atmospheric data; the [red]
dot-dash curves utilize K2K, MINOS, and CHOOZ data; the dashed [blue] curves
utilize all the data sets: atmospheric, K2K, MINOS, and CHOOZ.}
\label{fig1}
\end{figure}

\begin{figure}
\includegraphics*[width=3.3in]{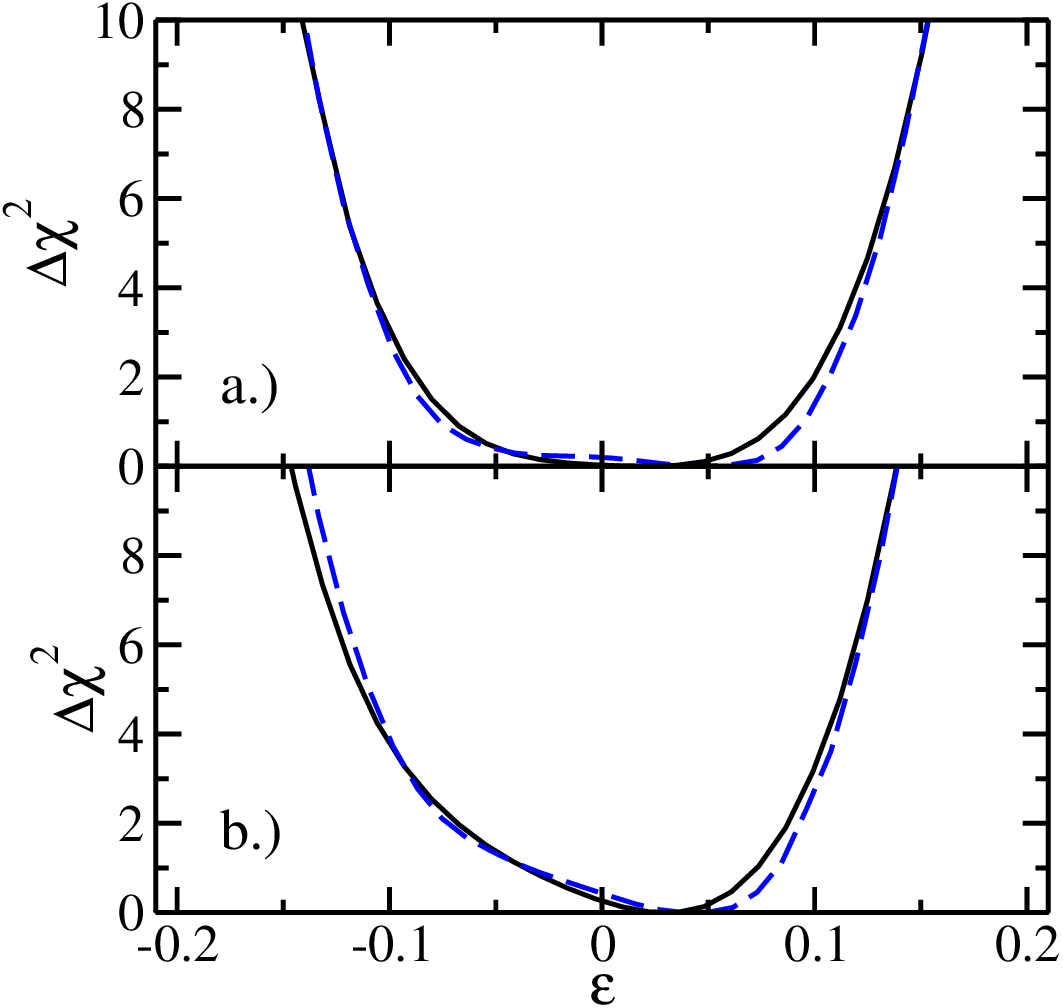} 
\caption{[color online] $\Delta\chi^2$ versus $\varepsilon$ for the a.)
sub-dominant approximation and b.) full three-neutrino calculation. The  [black]
solid curve utilizes only atmospheric data; the dashed [blue] curve utilizes all
the data sets: atmospheric, K2K, MINOS, and CHOOZ. }
\label{fig2}
\end{figure}

We begin by examining the mass-squared difference $\Delta_{32}$. We plot 
$\Delta\chi^2$ versus $\Delta_{32}$ using both the sub-dominant approximation, 
Fig.~1a, and the full three-neutrino calculation, Fig.~1b, with
$\theta_{13}$ and $\theta_{23}$ as varied parameters.  The [black] solid curves
are obtained from the Super-K atmospheric data alone. The [red]  dash-dot curves
employ the K2K, MINOS, and CHOOZ data, omitting the Super-K atmospheric data. 
These curves are largely determined by the recent MINOS data which constrain the
mass-squared difference more so than Super-K, as is well known. The analysis
utilizing all of the data sets  (atmospheric, K2K, MINOS, and CHOOZ) is depicted
by the [blue] dashed curves.  Notice that although the Super-K atmospheric data
is not as constraining as MINOS, it combines with MINOS to produce a reduced
bound, particularly from above.  In comparing the approximation, Fig.~1a, with
the full calculation, Fig.~1b, we see that the sub-dominant approximation is
useful for determining the mass-squared difference $\Delta_{32}$ . A very
careful inspection will reveal that the full three-neutrino analysis produces a
slightly larger bound than does the sub-dominant approximation. Our results are
$\Delta_{32}=0.25^{+0.02}_{-0.03}$ eV$^2$ at the 90\% confidence level.  (The
errors quoted for our calculations will be for $\Delta\chi^2=6.25$, the 90\%
confidence level for a three parameter fit.)

We next present $\Delta\chi^2$ versus $\varepsilon=\theta_{23}-\pi/4$ using the
sub-dominant approximation, Fig.~2a, and the full three-neutrino calculation, 
Fig.~2b, with $\theta_{13}$ and $\Delta_{32}$ as varied
parameters. We express our result  in terms of $\theta_{23}$, rather than
$\sin^22\,\theta_{23}$ or $\sin^2\theta_{23}$, because the oscillation
probabilities truly are a function of $\theta_{23}$. The [black] solid curve
again represents the Super-K atmospheric data alone. The [blue] dashed curve
represents the results from all data sets: Super-K atmospheric, K2K, MINOS, and
CHOOZ. Adding K2K, MINOS, and CHOOZ hardly alters the Super-K result. We do not
present the results for K2K, MINOS, and CHOOZ alone because this data does not
yield a reasonable constraint on $\theta_{23}$ when treated as a linear variable
with a varied $\theta_{13}$ included in the analysis. Only in a two neutrino
analysis does K2K and MINOS restrict the appropriate variable $\sin^2
2\theta_{23}$. Comparing the sub-dominant approximation with the full
calculation, we see that the full three-neutrino probabilities produce an
allowed region which is much more asymmetric about $\varepsilon=0$. In fact, we
find a statistically insignificant indication that $\theta_{23}$ is greater than
$\pi/4$, maximal mixing. The ability of atmospheric data to determine the octant
of $\theta_{23}$ has also been investigated in \cite{Choubey:2006}.
We find the value at 90\% CL is
$\varepsilon=0.03^{+0.09}_{-0.15}$.

\begin{figure}
\includegraphics*[width=3.3in]{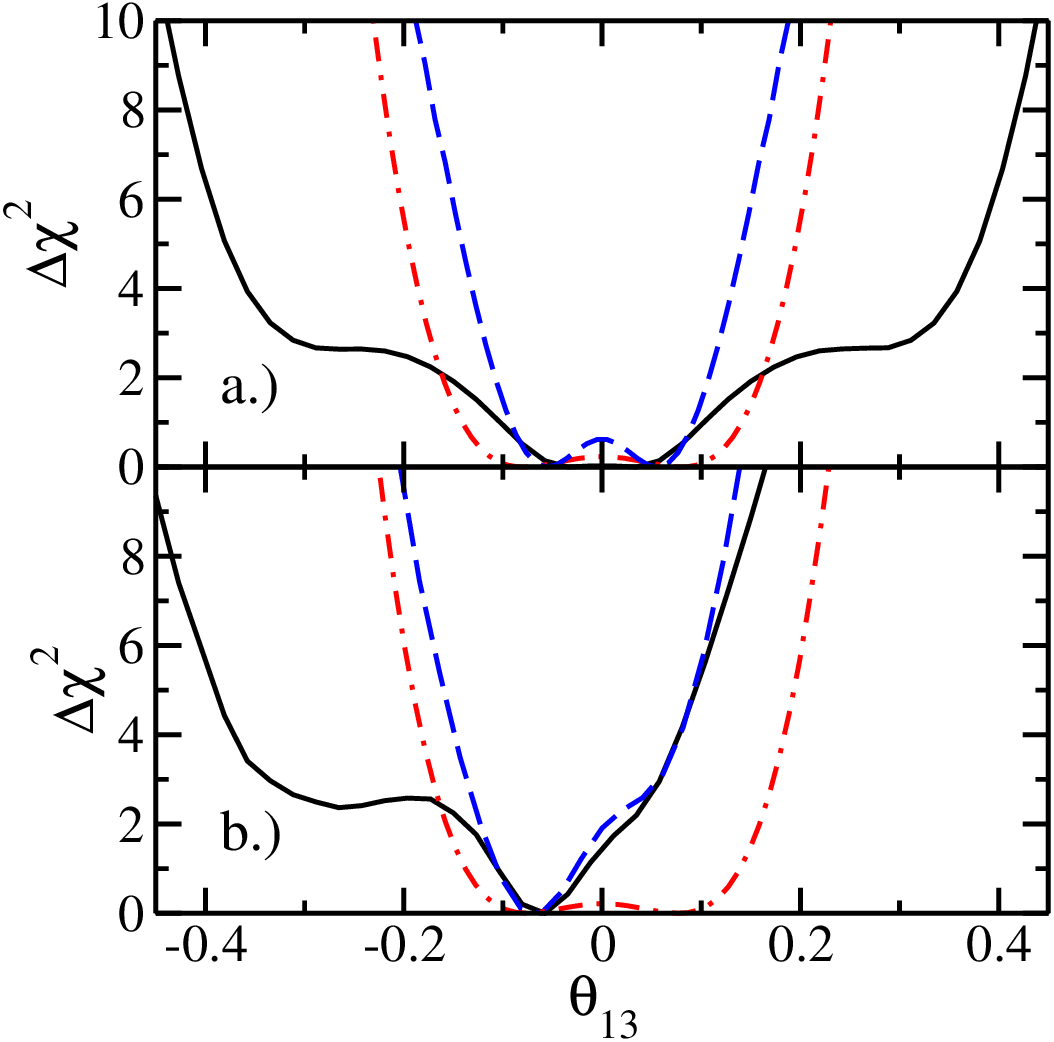} 
\caption{[color online]    The same as Fig.~\protect\ref{fig1} except
$\Delta\chi^2$ versus $\theta_{13}$ 
is presented.  }
\label{fig3}
\end{figure}

In Fig.~3a, we present $\Delta\chi^2$ versus $\theta_{13}$ calculated in
the sub-dominant approximation and full three-neutrino formulation with
$\theta_{23}$ and $\Delta_{32}$ as varied parameters. Previously, it has been
shown  \cite{skatm4} that in the sub-dominant approximation the atmospheric data
alone
restrict $\theta_{13}$.   Focusing upon our sub-dominant
calculation, Fig.~3a, the [black] solid  curve depicts the corresponding result
from our analysis.  As noted in the Appendix, our analysis quantitatively
reproduces
 the results in Ref.~\cite{skatm4}.  For $\Delta\chi^2 < 4.6$, the
90\% confidence level for a two-neutrino analysis, we both find
$\sin^2\theta_{13} < 0.14$ (or $\vert\theta_{13}\vert < 0.38$). This is a very
important calibration of our analysis tool.  The effect of $\theta_{13}$ on
atmospheric oscillations is small, and obtaining the same result implies we are
reproducing small effects, not just the global features of the analysis.  The
dash-dot [red] curve in Fig.~3a is the result of analyzing the K2K,
MINOS, and CHOOZ data, neglecting the Super-K atmospheric data. This curve is
mainly determined by the CHOOZ data. We see that CHOOZ is more constraining on
$\theta_{13}$ than is the Super-K atmospheric data. However, the dashed [blue]
curve presents the results utilizing all of the data sets and shows that the
Super-K data does somewhat reduce the error on $\theta_{13}$; this is due to the
indirect effect arising from Super-K further constraining the mass-squared
difference $\Delta_{23}$. To obtain the constraints on $\theta_{13}$ implied by
the Super-K atmospheric data, it is important to use the data from
Ref.~\cite{skatm4} which is more finely binned than earlier Super-K
\cite{skatm3} work. Note that the curves in the sub-dominant approximation are
symmetric about $\theta_{13}=0$ as is manifest from the approximate oscillation
formulae, Eq.~(\ref{subd}).   In
Refs.~\cite{Balantekin:2008,Fogli:2008,Maltoni:2008}, it has been observed that
recent data imply a statistically insignificant non-zero value for
$\theta_{13}$; our results are likewise consistent.

Turning to the full three-neutrino calculation, Fig.~3b, we find $\Delta \chi^2$
to be very asymmetric with a strong preference for negative $\theta_{13}$ when
using
only Super-K data, the [black] solid curve. The [red] dash-dot curve employs
only K2K, MINOS, and CHOOZ
data; it is symmetric about the origin so that the asymmetry present when all
data
is included, 
the [blue] dashed curve, is due to the Super-K data. What is more, we see the
novel result \cite{Escamilla:2008} that $\theta_{13}$ is constrained from above
by the Super-K atmospheric data, not by CHOOZ, while it is constrained from
below primarily by CHOOZ.

\begin{figure}
\includegraphics*[width=3.3in]{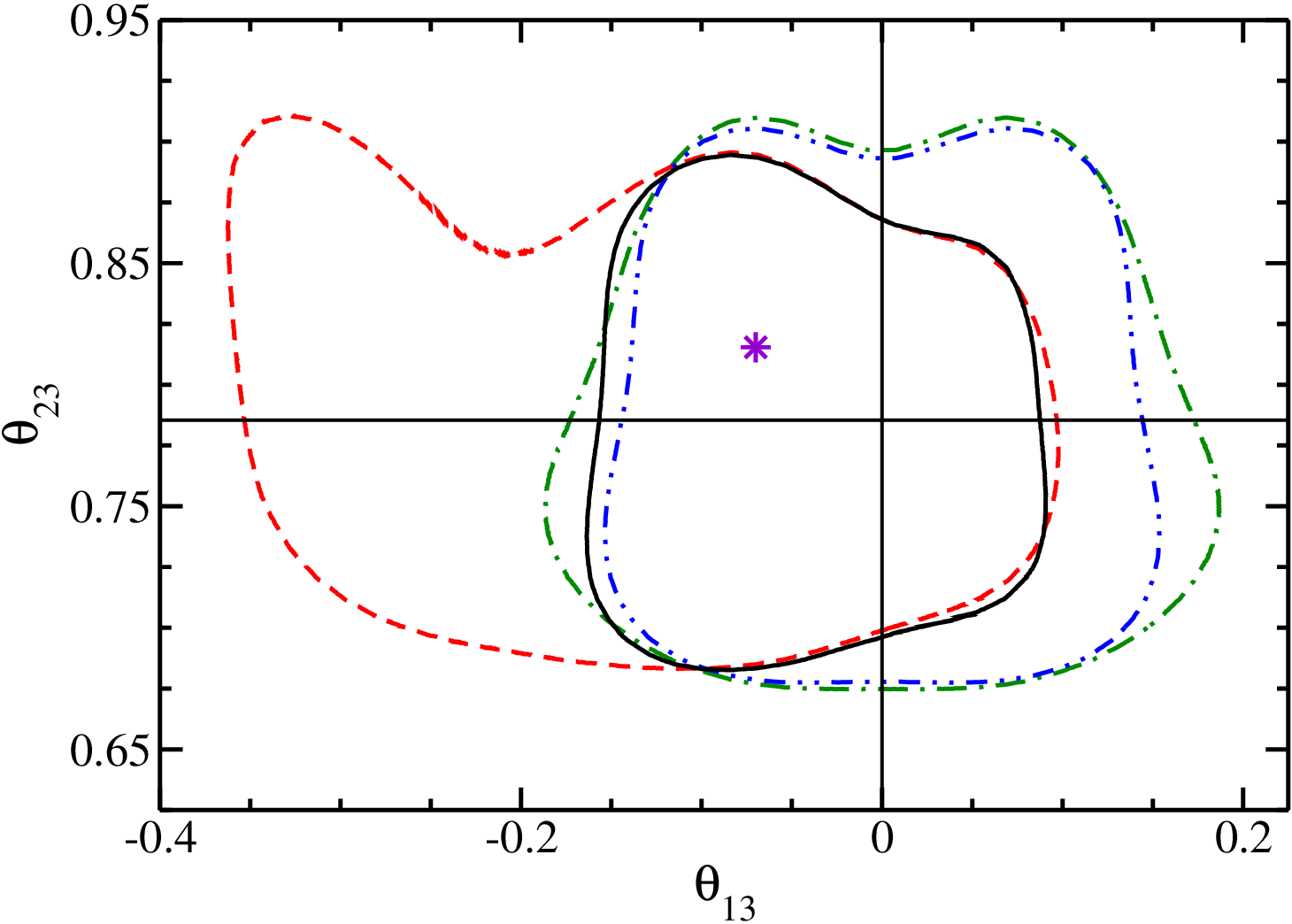}
\caption{[color online] The allowed region at 90\% CL for the parameters
$\theta_{13}$ 
and $\theta_{23}$.  The solid [black] curve uses the full three-neutrino
oscillation probabilities 
and  all the data sets: CHOOZ, K2K, MINOS, and Super-K atmospheric.  The
dash-dot-dot [blue] 
curve represents the use of the subdominant approximation and all the data 
sets.  The dashed [red] curve depicts the use of the full three-neutrino 
oscillation probabilities and only the Super-K data.  The dash-dot 
[green] curve depicts the use of the subdominant approximation and the 
Super-K data alone. The vertical (horizontal) straight line marks
$\theta_{13}=0.0$ ($\theta_{23}=\pi/4$). The [violet] star indicates the
location of the minimum for the analysis
that includes all the data sets and the use of the full three-neutrino
oscillation probabilities.}
\label{fig4}
\end{figure}

This conclusion is further reinforced by looking at the allowed region for the
parameters $\theta_{13}$ and $\theta_{23}$ as depicted in Fig.~\ref{fig4}. We
plot the 90\% confidence level of $\Delta\chi^2=4.61$ for a two parameter
analysis as we fix the third parameter $\Delta_{32}$ in calculating these
curves. The dash-dot [green] curve depicts the results for Super-K atmospheric
data alone in the sub-dominant approximation. We compare this with  the dashed
[red] curve which also utilizes only the Super-K atmospheric data alone but
incorporates the full three-neutrino probabilities.  Again, we see the
significant change brought about by incorporating the linear, and higher-order,
terms in $\theta_{13}$. The allowed region grows, favoring negative
$\theta_{13}$. The dash-dot-dot [blue] curve utilizes all the data in the
sub-dominant approximation. It is similar to the dash-dot [green] curve  because
the mass-squared difference $\Delta_{32}$ is fixed when calculating the curves;
the main effect of the MINOS experiment is to restrict $\Delta_{32}$. Finally,
the solid [black] curve utilizes all the data and the full three-neutrino
oscillation probabilities. Note that the upper bound on $\theta_{13}$ is similar
to that from the dashed [red] curve, i.e. the curve also utilizing the full
three-neutrino oscillation probabilities but only the Super-K data. For the lower
bound on $\theta_{13}$, however, we find similarities to the dash-dot-dot [blue]
curve where the restriction on $\theta_{13}$ originates primarily from the CHOOZ
experiment. Thus we again see that the upper bound on $\theta_{13}$ no longer arises 
from the CHOOZ experiment but is determined by the Super-K atmospheric experiment,
while the lower bound continues to come from the CHOOZ experiment. Our 
final result for this mixing angle is $\theta_{13}=-0.07^{+0.18}_{-0.11}$.

The principle effect of utilizing the full three neutrino oscillation
probabilities is the alteration of the shape of the allowed region for
$\theta_{13}$, particularly the introduction of the asymmetry about zero. The
absolute minimum for $\chi^2$ is lowered by only 1.3 \cite{Escamilla:2008}. This
is because the minima are very close to $\theta_{13}=0$ where the linear and
higher order terms contribute little.

\section{Discussion}

The most striking differences between the sub-dominant approximation and the
full three-neutrino probabilities were seen in the determination of the mixing
angle $\theta_{13}$, Fig.~\ref{fig3}.  Additionally, the deviation of
$\theta_{23}$ from maximal mixing also produced noticeable features, though less
striking, Fig.~\ref{fig2}.  Clearly, the two features are non-trivially linked
as demonstrated in the allowed regions depicted in Fig.~\ref{fig4}.  In fact, we
see from Fig.~\ref{fig3} that the Super-K data is the source of the asymmetry
about $\theta_{13}=0$ in the full three-neutrino model.  To flesh out which
subset of the Super-K data results in these asymmetries, we examine the various
contributions of the data to $\chi^2$ for a fixed positive and negative value of
the mixing angle $\theta_{13}$ taken to be $\pm 0.15$. The total difference in
$\Delta\chi^2$ for $\theta_{13}=+0.15$ and $\theta_{13}=-0.15$ is $\sim 7.0$.
Focusing on the fully contained events, we find that two thirds of this change
in $\Delta\chi^2$  between the positive and negative values of the mixing angle
comes from the sub-GeV electron-like events.  Half of the total change in
$\Delta\chi^2$ (3.5) arises from a single angular bin within this subset of
data, namely the bin for $e$-like events in which the detected charged lepton has
zenith angle
$\vartheta$ satisfying $-0.8<\cos\vartheta<-0.6$ and momentum less than 250 MeV.
The detected leptons in this bin are produced by neutrinos which travel along a
very long baseline upward through the earth.  Such neutrinos fall into the
region of $L/E$ where we have previously shown effects linear in $\theta_{13}$
to be significant in the oscillation probabilities $\mathcal{P}_{e \mu}$ and
$\mathcal{P}_{\mu \mu}$ \cite{Latimer:2005a,Latimer:2005b, Latimer:2005c}; such
effects occur in a region where the sub-dominant approximation is the leading
term in an expansion which is not convergent. Terms linear in
$\theta_{13}$ can be even more significant \cite{Peres:2009} should an
atmospheric oscillation experiment be able to take data at energies below 100
MeV.
We comment that the angular bin, $-1.0<\cos\vartheta< -0.8$, is also in this
$L/E$ region for the low-energy neutrinos; however, a typical neutrino which
produces leptons in this bin passes through the earth's higher density core.  We
find that the core suppresses the oscillations and thus this angular bin is not
as sensitive to the effects linear in $\theta_{13}$.

\begin{figure}
\includegraphics*[width=3.3in]{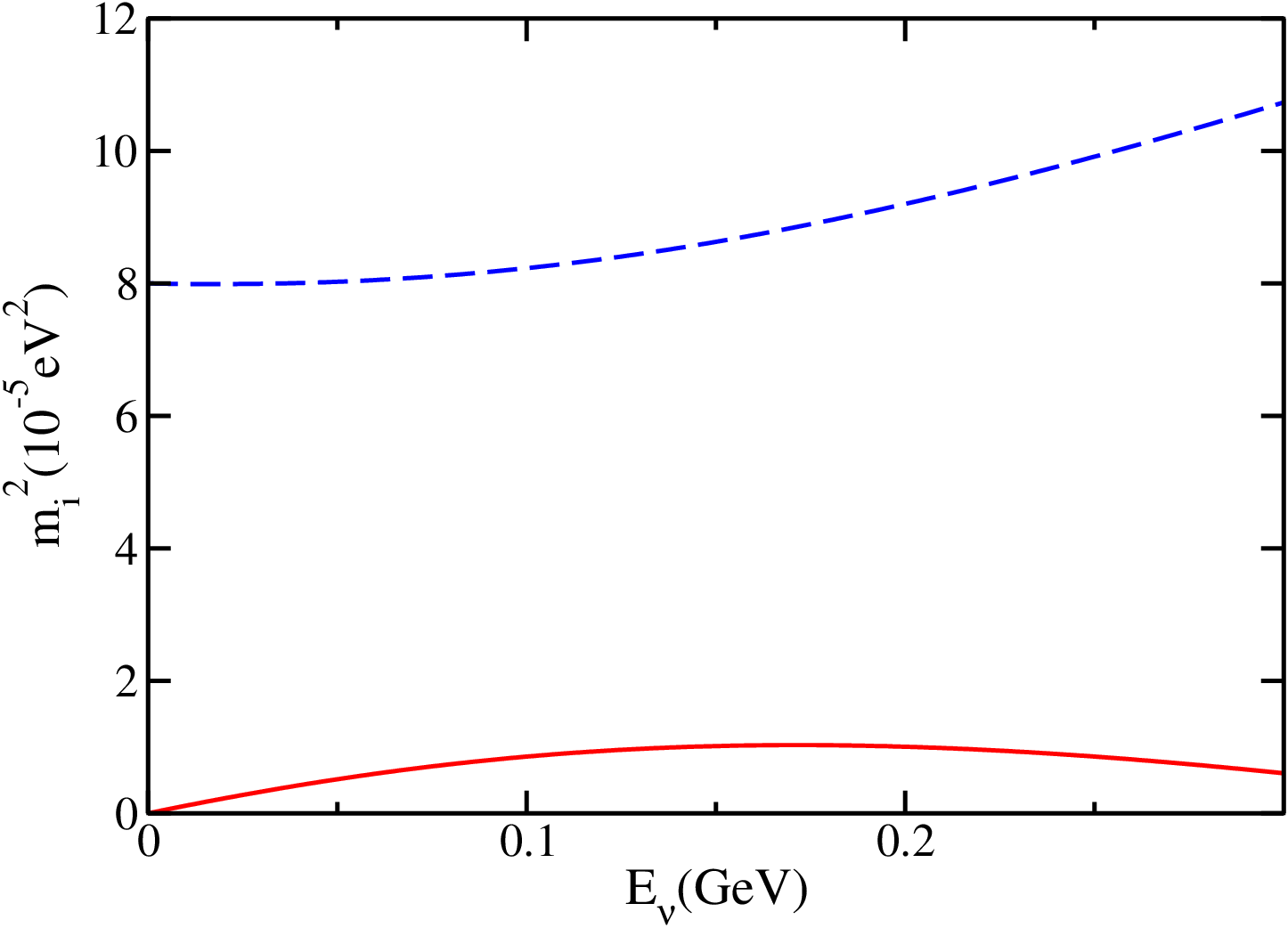}
\caption{[color online] The two lowest eigenvalues of the effective mass matrix
in a constant density mantle as a function of neutrino energy $E_\nu$. The [red]
solid curve is $(m_1)^2(E_\nu)$, and the [blue] dashed curve is
$(m_2)^2(E_\nu)$.} 
\label{fig5}
\end{figure}

The preference of the data for negative $\theta_{13}$ can be linked to the
excess of $e$-like events in the sub-GeV data set \cite{Peres:2004}. This excess
is not present in the $\mu$-like data or in the multi-GeV data so that an
overall renormalization of the atmospheric flux cannot account for the excess. 
To understand the role of the data in extracting $\theta_{13}$ and
$\theta_{23}$, we examine the relevant oscillation probabilities in the limit of
a constant density mantle and sub-GeV neutrino energies, keeping only terms
linear in $\theta_{13}$ and $\varepsilon$ and averaging over the $\Delta_{32}$
oscillations; these approximations have been discussed previously
\cite{Peres:2004,Latimer:2005c}.  As detailed in the Appendix, the electron-like
events at the Super-K detector are related to the $\nu_e$ survival probability
and the $\nu_\mu$ conversion probability via $\mathcal{R}_e={\mathcal
P}_{ee}+r{\mathcal P}_{e\mu}$ where $r$ is the ratio of the $\nu_\mu$ to $\nu_e$
flux at the source. This yields the approximate expression
\begin{eqnarray}
&&\mathcal{R}_e \approx 1 \nonumber\\
&&~+ r\sin^2 2\theta_{12}^m \left[ \frac{1}{2} - \frac{1}{r} + \cot (2
\theta_{12}^m)\, \theta_{13} - \varepsilon \right] \sin^2 \varphi_{21}^m.~~~~~~
\end{eqnarray}
Here, $\theta_{12}^m$ is the effective mixing angle in matter;
additionally, the phase $\varphi_{21}^m$ employs the effective mass-squared
difference in matter corresponding to $\Delta_{21}$.  In this approximation, we
can understand how to effect an excess of electron-like events for sub-GeV
neutrinos over a long baseline,
\begin{equation}
\frac{1}{2} - \frac{1}{r} + \cot (2 \theta_{12}^m) \,\theta_{13} - \varepsilon >
0. \label{ineq}
\end{equation} 
Using the same approximations, we can simply express the MSW resonant energy
\begin{equation}
 E_R=\frac{\Delta_{21}\cos2\theta_{12}}{2V\cos^2\theta_{13}}\,,
\end{equation} 
with $V \approx 1.7 \times 10^{-13}$ eV in the mantle; this yields $E_R$ on the
order of 100 MeV.  
This resonance is apparent when we plot the eigenvalues of the effective
mass-squared matrix in the mantle, Fig. \ref{fig5};  the ``resonance'' is
indicated by the slight bowing in the curves toward each other and is located at
the point where the effective mass-squared difference is minimal.

At the resonant energy, one has $\theta_{12}^m =  \pi/4$; for neutrino energies
above the resonance, the effective mixing angle in matter increases up to
$\pi/2$.  As a consequence, for neutrino energies above 100 MeV in the mantle,
the function $\cot (2 \theta_{12}^m)$ is negative; to reiterate, the coefficient
of the $\theta_{13}$ term in the inequality, Eq.~(\ref{ineq}), is negative.  We
note that for these low energy atmospheric neutrinos $r \sim 2$ so that the
first two terms of the inequality approximately sum to zero.  If $\theta_{13}$
is restricted to positive values, then the mixing angle $\theta_{23}$ must lie
in the first octant ($\varepsilon <0$) in order to account for the excess in
$\mathcal{R}_e$.  However, if we allow $\theta_{13}$ to run the full range of
allowed parameter space in a CP conserving theory, then a negative value of this
mixing angle can easily accommodate the excess in $\mathcal{R}_e$, even
permitting $\theta_{23}$ to lie in the second octant as is the case in our
analysis.

\begin{figure}
\includegraphics*[width=3.3in]{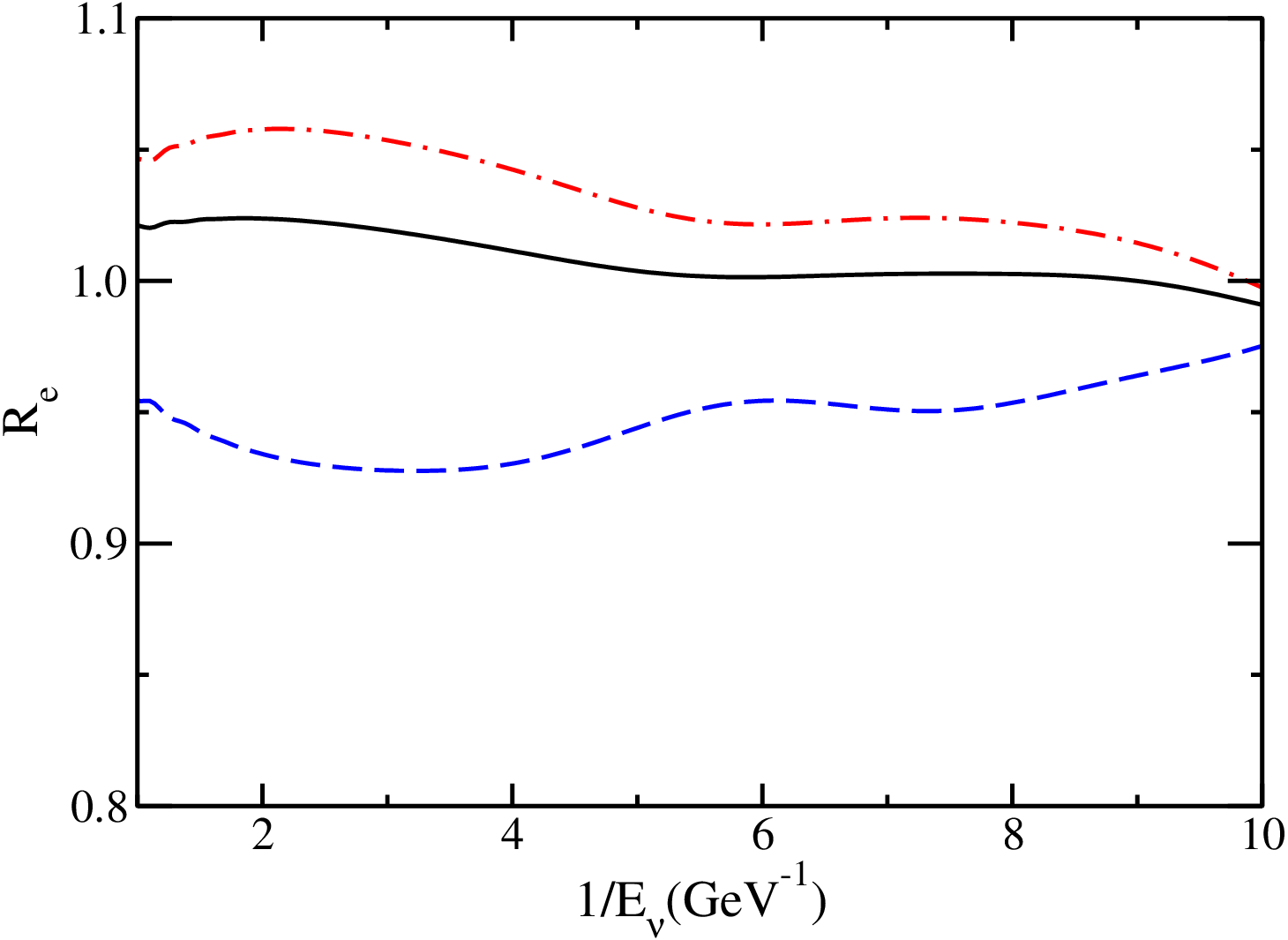}
\caption{[color online] The experimentally measured quantity $\mathcal{R}_e$
versus the
inverse neutrino energy $E_\nu^{-1}$ in the sub-GeV region for angular bin
$-0.8<\cos\vartheta<-0.6$. The solid [black] curve is the result for the best
fit parameters, the dash-dot [red] curve is for $\theta_{13}=-0.15$ and the
dashed [blue] curve is the result for $\theta_{13}=+0.15$.}
\label{fig6}
\end{figure}

To demonstrate the point regarding the effect of terms linear in $\theta_{13}$,
we plot $\mathcal{R}_e$ in Fig.~\ref{fig6} for sub-GeV neutrinos in angular bin
$-0.8<\cos\vartheta<-0.6$. The solid [black] curve employs our best fit
parameters.  To show the effect of $\theta_{13}$, we also plot the $e$-like events
for $\theta_{13}=\pm 0.15$ with $\Delta_{23}$ and $\theta_{23}$ unchanged.
The dash-dot [red] curve has negative $\theta_{13}$, and the dashed [blue] curve
has positive $\theta_{13}$.  It is clear that a negative value of this mixing
angle permits an excess of $e$-like events for sub-GeV neutrinos.

Returning to Fig.~\ref{fig4}, we see how the full three-neutrino oscillation
probabilities jointly affect the allowed region for the two mixing angles
$\theta_{23}$ and $\theta_{13}$.  Using the subdominant approximation, the
dash-dot-dot [blue] curve represents the allowed region when all the data is
employed.  As expected, the region is symmetric about $\theta_{13} =0$, and we
note that the actual value of $\theta_{13}$ has little impact upon the allowed
values of $\theta_{23}$, save the neighborhood immediately around
$\theta_{13}=0$.  Inclusion of the higher order terms in the oscillation
probability dramatically alters this picture.  As discussed above, the data now
favor negative $\theta_{13}$, with the atmospheric Super-K data shrinking the 90\% 
CL contour, the solid [black] curve, for positive $\theta_{13}$.  No longer is
the contour symmetric about a particular value of $\theta_{13}$; hence, the true
value of this mixing angle will impact the allowed region for the $\theta_{23}$
mixing angle.  In particular, the allowed region for $\theta_{23}$ shrinks as
$\theta_{13}$ approaches positive values.   In the future, should a reactor
neutrino experiment confirm a nonzero value for $\vert \theta_{13}\vert$, it
will have interesting consequences for the allowed value of $\theta_{23}$.  With
such a measurement, we would perhaps see two true local minima in the $\Delta
\chi^2$ versus $\theta_{13}$ plot in Fig.~\ref{fig3}.  The impact upon
Fig.~\ref{fig4} would be to separate the jointly allowed regions into two
disconnected curves with the limit on $\theta_{23}$ more tightly constrained for
positive values of $\theta_{13}$.

\section{Conclusions}

As we enter into the era of precision neutrino experiments, small effects, such
as those arising from $\theta_{13}$ or the octant of $\theta_{23}$, require a
careful treatment in the analysis. Future reactor experiments
\cite{Guo:2007,dchooz,reno} are sensitive to $\theta_{13}^2$ and thus can
determine the magnitude of $\theta_{13}$, but not its sign. Long-baseline
experiments, e.g., \cite{t2k}, will contain small effects that are linear in
$\theta_{13}$, while an upgraded Super-K will produce additional data in the
region where we have found significant effects linear in $\theta_{13}$. How
these different data interplay with each other in determining $\theta_{13}$,
including its sign, and the octant of $\theta_{23}$ will be most interesting.

In Ref.~\cite{dharam}, it was shown that the constancy of $R_e$ imposes an upper
bound on $\vert \theta_{13} \vert$ as well as constrains $\theta_{23}$ to be
near maximal mixing.
We find that present atmospheric data restrict the value of $\theta_{13}$ from
above, while the limit from below remains as determined by CHOOZ. We find
$\theta_{13}=-0.07^{+0.18}_{-0.11}$, assuming no CP violation. We have
investigated which data points lead to the asymmetry in $\theta_{13}$ and find
that it is the atmospheric data in the very long baseline region previously
noted \cite{Latimer:2005a,Latimer:2005b, Latimer:2005c} to have significant
terms linear in $\theta_{13}$. We further found that the Earth's MSW effect
plays an
important role as it increases the effective value of $\theta_{12}^m$ in matter
such that the atmospheric data provides a strict upper bound on $\theta_{13}$. 
Further, the data producing the preference for a negative $\theta_{13}$ is data
with an excess of $e$-like events, ${\mathcal R}_e>1$. Allowing $\theta_{13}$ to
be
negative supports this excess and permits $\theta_{23}$ to be in the second
octant. The parameters $\theta_{13}$ and $\theta_{23}$ are found to be
correlated; the statistically insignificant negative value for the minimum of
$\theta_{13}$ relates to the minimum for $\theta_{23}$ being statistically
insignificantly in the second octant, and the error in $\theta_{23}$ is dependent
on the value of $\theta_{13}$. Future measurements of $\theta_{13}$ will impact 
the allowed value for $\theta_{23}$.

For $\Delta_{32}$ and $\theta_{23}$ we find $\Delta_{32}=0.25^{+0.02}_{-0.03}$
eV$^2$ and $\theta_{23}-\pi/4=0.03^{+0.09}_{-0.15}$, where the use of the full
three-neutrino oscillation probabilities leads to the asymmetry in the errors. We
find that a quantitative analysis requires utilizing the more finely binned
atmospheric data of Ref.~\cite{skatm4}, the use of the full three-neutrino
oscillation probabilities, and the inclusion of the full three-neutrino MSW
effect.

\appendix*
\section{Experimental simulation}

In this appendix, we present the computational tools we use to analyze the
Super-K atmospheric, CHOOZ, K2K, and MINOS experiments. The analysis tool for
the Super-K atmospheric data is similar to that being used by others
\cite{schwetz:2008,Gonzalez:2008,Fogli:2006}; however, it is distinct in that 
we employ a full three-neutrino oscillation probability rather than an
approximate expansion, use a full three-neutrino treatment of the Earth's MSW
effect, and include a model of the multi-ring data. The analysis of CHOOZ, K2K,
and MINOS data is standard. Additional details can be found in
Ref.~\cite{jesus}.  Also, in this appendix, we demonstrate the efficacy of our
analysis tools by comparing our results with others' when appropriate.

The appendix is organized as follows.
We first discuss the Super-K atmospheric experiment, beginning with the
contained events followed by the upgpoing muon events.  Then we discuss our
statistical treatment of this experiment. Finally, we include a similar
discussion for the CHOOZ, K2K, and MINOS experiments.

\subsection{Super-K contained events}

In order to observe atmospheric neutrinos at Super-Kamiokande, the neutrinos
must interact with matter in either the detector or the surrounding environ to
produced charged particles.  The direction and energy of these charged particles
can be deduced from the Cherenkov light they emit while traveling through the
water-filled detector; from this data, one can infer on average the direction
and energy of the initial neutrino.  The Super-K experiment classifies the
various detections in terms of the production point of the charged lepton, the
number of charged particles produced,  and their subsequent motion through the
detector.

Contained events refer to events in which the charged lepton is produced by the
neutrino within the detector.  These events are subdivided into  fully contained
and partially contained events.   If an event is fully contained, then the
charged lepton(s) produced within the detector do not escape the detector.  An
event is partially contained if the charged lepton(s) exit the detector. 
Finally, these two data sets are  further separated into single-ring and
multi-ring events according to the number of charged particles produced by the
neutrino; if only one charged lepton is observed in the detector, it is termed a
single-ring event. We first discuss the  fully contained single-ring events and
then extend this analysis to the other classes of data.
The fully contained  single-ring events are statistically the most significant
subset of the data and the cleanest to analyze. Preliminary discussions of the
analysis technique utilized for the fully contained events can be found in
Refs.~\cite{jesus:2007,jesus:2008}.

The Super-K detector distinguishes between electrons and muons by the fuzziness
of the Cherenkov ring generated by the charged lepton; however, the detector
cannot differentiate an electron $e^-$ from a positron $e^+$ or a $\mu^-$ from a
$\mu^+$.  Given flavor and charge conservation, the detector can only determine
if an event is $e$-like, originating from either a $\nu_e$ or $\overline{\nu}_e$
interaction, or $\mu$-like, originating from either a $\nu_\mu$ or
$\overline{\nu}_\mu$ interaction.  
 Thus the detector counts charged leptons of flavor $\alpha$ in energy bin $m$
and zenith angular bin $n$ over the run time $T$

\begin{equation}
N_\alpha^{nm}=\sum_{\nu,\overline\nu}\left(
\frac{\mathrm{d}N_{\alpha\rightarrow\alpha}^{nm}}{\mathrm{d}t}+\frac{\mathrm{d}
N_{\beta\rightarrow\alpha}^{nm}}{\mathrm{d}t}\right)T\,\,,
\label{oscfin}
 \end{equation}
where the quantity $ \mathrm{d} N_{\beta\rightarrow\alpha}^{nm}/\mathrm{d} t$
represents the rate at which a neutrino, created in the atmosphere with flavor
$\beta$, will be detected as an $\alpha$-like event in the appropriate energy and
angular bins within the detector.

This rate depends upon the atmospheric neutrino flux, the neutrino oscillation
probability from source to detector, the kinematics of the charged lepton
production, and the detector efficiencies.  We may write it as
\begin{widetext}
\begin{eqnarray}
\frac{\mathrm{d}N_{\alpha\rightarrow\beta}^{nm}}{\mathrm{d}
t}&=&N\,\int_{E_{vis}^{m,min}}^{E_{vis}^{m,max}} \mathrm{d}E_{vis} \int
\mathrm{d} \cos \theta_\nu \int \,\mathrm{d}E_\nu
\int \mathrm{d}\cos \theta_s \int \mathrm{d}\phi_s\nonumber\\
&\times&\varepsilon (E_{vis} )\,\frac{\mathrm{d}^2\,\Phi_\alpha
(E_\nu,\cos\theta_\nu)}{\mathrm{d}E_\nu\,\mathrm{d}\cos\theta_\nu}\,
\tilde{\mathcal P}_{\alpha\beta}(E_\nu,\,cos\theta_\nu )\,
\frac{\mathrm{d}^3 \sigma_\beta (E_\nu; \,\,E_\ell, \,\cos \theta_s
)}{\mathrm{d}E_\ell\, \mathrm{d}\cos \theta_s \,\mathrm{d}\phi_s}\nonumber\\
&\times& \Theta(\cos\vartheta^{n,max}-\cos\vartheta)\,\,\Theta(\cos\vartheta-
\cos\vartheta^{n,min}) \,.
\label{oscbin}
\end{eqnarray}
\end{widetext}
We define the variables in Eq.~(\ref{oscbin}). $N$ represents the number of
target protons.  $E_{vis}$ is the energy measured by the detector (this quantity
is defined differently depending on the data sample); $E_{vis}^{m,max}$
($E_{vis}^{m,min}$) is the maximum (minimum) value of $E_{vis}$ for bin $m$.
(For single-ring fully contained events, $E_{vis}$ is simply the energy of the
created lepton, $E_{vis}=E_\ell$.)  $\vartheta$ is the zenith angle of the
detected charged lepton with $\cos\vartheta=1$ indicating the vertically
downward direction.
The relative angle between the incident neutrino and the produced charged lepton
are described by the $\theta_s$ scattering angle and the $\phi_s$ azimuthal
angle.
The energy of the incident neutrino is $E_\nu$ with zenith angle $\theta_\nu$.
The azimuthally averaged atmospheric neutrino flux for a neutrino of flavor
$\alpha$  is $\mathrm{d}^2\,\Phi_\alpha
(E_\nu,\cos\theta_\nu)/\mathrm{d}E_\nu\,\mathrm{d}\cos\theta_\nu$ which we take
from Ref.~\cite{Honda3d}.   $\varepsilon (E_{vis} )$ corresponds  to the
detection efficiency. $\mathrm{d}^{3}\sigma_\beta (E_\nu ; \,E_\ell, \,\cos
\theta_s )/\mathrm{d}E_\ell\, \mathrm{d}\cos \theta_s \,\mathrm{d}\phi_s$ is the
differential cross section for a neutrino of energy $E_\nu$ and flavor $\beta$
to produce a charged lepton of flavor $\beta$ with energy $E_\ell$ through a
scattering angle $\theta_s$. Although the differential cross section which
occurs in Eq.~\ref{oscbin} does not depend on the azimuthal angle $\phi_s$, the
geometry that determines in which angular bin an event lies does depend on
$\phi_s$. This is because the zenith angle $\vartheta$ of the charged lepton is
given in terms of the neutrino zenith angle $\theta_\nu$ and the scattering
angles $\theta_s$ and $\phi_s$ by
\begin{equation} 
\cos\vartheta=\cos\theta_s\,\cos\theta_\nu
-\sin\theta_s\,\sin\theta_\nu\,\cos\phi_s\,\,.
\label{angles}
\end{equation}

The energy range for atmospheric neutrinos as measured at Super-K requires the
use of several cross sections. At low energies, below 1 GeV, the dominant
process is charged-current quasi-elastic  scattering from the proton and the
nucleons in the oxygen nucleus in H$_2$O, e.g., $\overline\nu_e+p^+\rightarrow
e^++n$ and $\nu_e+n\rightarrow e^-+p^+$. At intermediate energies, peaking
around 1.5 GeV, the dominant process is single-pion resonance production, i.e.
$\nu_\alpha + N\rightarrow \ell_\alpha + N^*$ followed by $N^*\rightarrow
N'+\pi$. At higher energies, starting at 1 GeV and dominating above 10 GeV,
deeply inelastic scattering occurs, $\nu_\alpha+N\rightarrow \ell_\alpha +X$ 
We utilize the same set of cross sections as was used in Ref.~\cite{skatm3}.
The Heaviside functions are
inserted into Eq.~(\ref{oscbin}) in order to restrict the values of
$\theta_\nu$, $\theta_s$, and $\phi_s$ to values which produce a value for
$\vartheta$ that lies within bin $n$.

The incident neutrino's zenith angle $\theta_\nu$ does not uniquely determine
the path length $L$, as  neutrinos are produced at a variety of vertical heights
$h$ in the atmosphere. We thus introduce the oscillation probability
$\tilde{\mathcal P}_{\alpha\beta}(E_\nu,\cos\theta_\nu )$ averaged over this
production height in the atmosphere 
\begin{eqnarray}
&&\tilde{\mathcal P}_{\alpha\beta}(E_\nu,\cos \theta_\nu)= \nonumber \\
&&~~\int_0^\infty \mathrm{d}h\, {\rm P}_\alpha(h, E_\nu)\,
{\mathcal P}_{\alpha \beta}(L(h,\cos\theta_\nu)/E_\nu)\,,
\label{phat}
\end{eqnarray}
where $L$ is related to $h$ and $\cos\theta_\nu$ by
\begin{equation}
L=\sqrt{R^2 \,\cos^2\theta_\nu + h\,(2R+h)}-R\,\cos\theta_\nu\,\,,
\end{equation}
with $R$ the radius of the Earth. ${\rm P}_\alpha (h, E_\nu)$ is the normalized
probability for a neutrino of flavor $\alpha$ to be created at a height $h$, a
quantity we take from Ref.~\cite{slant}.

The neutrino oscillation probability in vacuum ${\mathcal P}_{\alpha\beta}
(L/E_\nu)$ is given in Eq.~(\ref{exact}); however, the coherent
forward scattering of neutrinos on matter alters the probability for those
neutrinos which pass through the Earth  \cite{msw1,msw2}.   Neutral current
interactions between the neutrinos and matter are not flavor dependent leaving
the oscillation probabilities unaffected; however, charged current interactions
will introduce into the Hamiltonian a flavor dependent potential.  In the flavor
basis, we may write \cite{Hannabuss:2000} the neutrino evolution equation as
\begin{equation}
i \partial_t \nu_f = \left[ \frac{1}{2E_\nu} U \mathcal{M} U^\dagger +
\mathcal{V} \right]\nu_f \label{mswev}
\end{equation}
where $\mathcal{M}= \mathrm{diag}\, (0,\Delta_{21},\Delta_{31})$ and the
operator $\mathcal{V}$ operates on only the electron flavor with a
magnitude $\mathcal{V}= \sqrt{2} G_F N_e(x)$ in which $N_e(x)$ is the electron
number density of the matter.
We note that for anti-neutrinos, this potential has the opposite algebraic sign.
In matter of constant density, one may diagonalize the Hamiltonian to determine
the effective mass-squared differences in a new propagation basis; this basis is
related to the flavor states by a modified effective mixing matrix in the
matter.   The density of the earth may be approximated as piecewise constant
\cite{prem} so that neutrinos which travel through the earth may traverse
regions of varying density.
For certain energies and mixing angles, neutrinos can undergo parametric
resonances when passing through regions of different densities
\cite{Akhmedov:1998}. 
For a piecewise constant density profile, we may use the the methods of
Refs.~\cite{Ohlsson:1999,Ohlsson:2000}  to exactly determine the neutrino
oscillation probability.  This method is computationally efficient as it merely
involves finding the local effective mass-squared differences and mixing angles,
and its exactness accurately accounts for effects which depend on small
parameters or effects due to parametric resonances.

The full calculation represented by Eq.~(\ref{oscbin}) is a
numerically intensive five-dimensional integration.  We desire an analysis tool
which is both numerically accurate and sufficiently computationally efficient
that we can scan a large swath of parameter space in a reasonable amount of
time.  To effect this, we make approximations in Eq.~(\ref{oscbin}) regarding
the scattering angle and efficiency terms.  First, we fix the scattering angle
$\theta_s$ to its average value $\overline\theta_s$ as a function of
$E_{vis}=E_\ell$ for each energy and angular bin as calculated by the Monte
Carlo simulations in Ref.~\cite{Messier}.  This eliminates the integral over
$\cos \theta_s$.  We are justified in doing so because, if this
integral is done last, the integrand 
is quite smooth and nearly linear over each bin, particularly for the finer
binning of the most recent data.
Thus, we expect this approximation to be quantitatively accurate, and this is
what we have found.  With this approximation, Eq.~(\ref{oscbin}) becomes

\begin{widetext}
\begin{eqnarray}
\frac{\mathrm{d} N_{\alpha\rightarrow\beta}^{nm}}{\mathrm{d}
t}&=&N\,\int_0^\infty \mathrm{d} E_\nu
\,\int_{E_{vis}^{m,min}}^{E_{vis}^{m,max}} \mathrm{d} E_{vis} \int \mathrm{d}
\cos \theta_\nu 
\int \mathrm{d} \phi_s\nonumber\\
&\times&\varepsilon (E_{vis} )\,\frac{\mathrm{d}^2\,\Phi_\alpha
(E_\nu,\cos\theta_\nu)}{\mathrm{d} E_\nu\,\mathrm{d} \cos\theta_\nu}\,
\tilde{\mathcal P}_{\alpha\beta}(E_\nu,\cos\theta_\nu )\,
\frac{\mathrm{d}^2 \sigma_\beta (E_\nu ;\, E_\ell, \cos \overline\theta_s
(E_{vis}))}{\mathrm{d} E_\ell\,\mathrm{d}
\phi_s}\nonumber\\
&\times& \Theta(\cos\vartheta^{n,max}-\cos\vartheta)\,\,\Theta(\cos\vartheta-
\cos\vartheta^{n,min}) \,,
\label{oscbin2}
\end{eqnarray}
\end{widetext}
where the differential cross section and the angle $\vartheta$ have been evaluated 
at $\cos(\theta_s)=\cos(\overline\theta_s)$.

Finally, we must determine the detector efficiency $\varepsilon(E_{vis})$, which
has not been furnished by the experimentalists. We can, however, extract it from
information provided. For no oscillations, Eq.~(\ref{oscbin})
becomes, utilizing our assumptions,
\begin{widetext}
\begin{eqnarray}
\frac {\mathrm{d} N_{\alpha}^{m}}{\mathrm{d}t}&=& \overline{\varepsilon}
(m)\,N\,\int_0^\infty \mathrm{d}E_\nu\,\int_{E_{vis}^{m,min}}^{E_{vis}^{m,max}}
\mathrm{d}E_{vis}   \int \mathrm{d} \cos \theta_\nu 
\int \mathrm{d} \phi_s\,\frac{\mathrm{d}\,\Phi_\alpha
(E_\nu,\cos\theta_\nu)}{\mathrm{d}E_\nu\mathrm{d}\cos\theta_\nu}\,
\frac{\mathrm{d}^3 \sigma_\beta (E_\nu ;\, E_\ell, \cos \overline\theta_s
(E_{vis}))}{\mathrm{d} E_\ell\, \mathrm{d} \cos \theta_s \,\mathrm{d} \phi_s}
\,\,\,
\label{oscbin3}
\end{eqnarray}
\end{widetext}
where we have assumed that the efficiency is primarily dependent on the 
lepton energy, $\varepsilon (E_{vis} )$. Further taking it as a constant
 $\overline{\varepsilon} (m)$ over each energy bin $m$, we pull it out of the 
integral. The Monte Carlo calculation of ${\mathrm{d}
N_{\alpha}^{m}}/{\mathrm{d} t}$ appears in Ref.~\cite{Ishitsuka}.  Performing
the integrals on the right hand side of Eq.~(\ref{oscbin3}), we can use the
Monte Carlo result to determine the average efficiency $\overline{\varepsilon}
(m)$  for each energy  bin.   This average is then used in Eq.~(\ref{oscbin2}).

To conclude the discussion of the single-ring fully contained analysis, we note
that there are ten angular bins of equal size in $\cos\vartheta$, going from
$+1$ (downward) to $-1$ (upward); ten energy bins for $e$-like events; and 8
energy bins for $\mu$-like events.  In total, this subset of the data consists
of 180 data points.

Within the fully contained data set, we also  have  multi-ring events. This data
correspond to neutrinos which interact inside the detector to produce more than 
one detected particle in the  final state.  We can calculate the multi-ring
event rate in a manner similar to the single-ring event rate provided we modify
the visible energy and neutrino event direction. In a  multi-ring event,  the
visible energy depends on  the number of  particles in the  final state, their
momenta,  and their scattering angles. There is not a simple and reliable way of
determining $E_{vis}$; as such, we must make some approximations. We use the
results of Monte Carlo simulations in Ref.~\cite{Messier} to estimate  the
average value of $E_{\nu}/E_{vis}$  for  each energy bin.   To determine the
angular distribution of these events, we assign a single scattering angle to the
final state particles.  To make a simple estimate of the average multi-ring
scattering angle for each energy bin $m$, $\overline\theta_{s,multi}(m)$, we fit our
no-oscillation event rate to the analogous no-oscillation  Monte Carlo
calculation from Ref.~\cite{Ishitsuka}.  Our analysis then uses
Eqs.~(\ref{oscfin},\ref{oscbin},\ref{angles}) to calculate the general event rate.

The next data set consists of the partially-contained events. In these, the
charged
lepton is created within the detector but has sufficient energy to escape the
inner detector and be detected by the outer detector (OD). If the energy
deposited in the OD appears to be less than that  needed for a muon to traverse
the OD, the event is classified as an OD stopped event; those depositing more
energy are termed OD through-going. We alter the definition of the visible
energy
\begin{equation}
E_{vis}=E_{inner}+E_{dead}+E_{anti}\,,
\label{evis3}
\end{equation}
\noindent
with $E_{inner}$ the total energy of the charged particle observed in the inner
detector, 
$E_{dead}$ the energy deposited in the region  between the inner detector and
the outer detector, and $E_{anti}$ the energy observed in the outer detector.  
The charged lepton energy is then a function of this visible energy; we set
$E_\ell(E_{vis})$ to its average value as taken from the Monte Carlo
calculations from Ref.~\cite{Messier}. The efficiencies are similarly taken for
each data bin from the same Monte Carlo no-oscillation calculations as described
above.
For these single-ring partially contained events, the same ten angular bins are
used, and there are four energy bins each for both the OD stopped and
through-going events, giving a total of 80 data points.

\subsection {Upgoing muons}

A second major subset of the atmospheric data arises from detecting muons
created by atmospheric neutrinos and anti-neutrinos interacting with the rock
surrounding the detector. These muons can then either stop in the detector,
called ``stopping muons'', or pass through the detector, called ``through-going
muons''. Typically the stopping muons have an $\mathcal{O}(\text{10 GeV})$
energy while the through-going muons have an $\mathcal{O}(\text{100 GeV})$
energy. Although statistically not as significant
as the contained events, these data provide important information in this higher
energy range.

Our analysis of these events exactly follows that given in
Refs.~\cite{Fornengo:1999,GonzalezGarcia:2000}. We do not include the effects of
the muon energy fluctuations as has recently been done in
Ref.~\cite{Gonzalez:2008}. The detection rate for stopping muons,
$S$, and through-going muons, $T$, is given by

\begin{widetext}
\begin{eqnarray} 
\frac{\mathrm{d} N_{\mu,S,T}^n}{\mathrm{d}t}&=&N_A\,\int_0^\infty
\mathrm{d}E_\nu\int_{\theta^{n,min}}^{\theta^{n,max}} \mathrm{d}\cos\theta_\nu
\int_0^{E_\nu} \mathrm{d}E_\mu \nonumber\\
&\times&\frac{\mathrm{d}^2\,\Phi_\alpha
(E_\nu,\cos\theta_\nu)}{\mathrm{d}E_\nu\,\mathrm{d}\cos\theta_\nu}\,\tilde{
\mathcal P}_{\alpha\mu}(E_\nu,\,\cos\theta_\nu )\,
\frac{\mathrm{d} \sigma_\mu (E_\nu; \,E_\mu)}{\mathrm{d} E_\mu} {\mathcal
R}(E_\mu,E_{th})
\,{A_{S,T}(E_{th}, \cos\theta_\nu)}\,.
\label{upmu}
\end{eqnarray}
\end{widetext}
We have assumed, appropriate for these high energies, that the scattering is
forward. This allows us to replace the charged lepton angle $\vartheta$ with 
the neutrino angle $\theta_\nu$; we may also perform the integration over the
scattering
angles $\theta_s$ and $\phi_s$ in the cross section. $N_A$ is Avogadro's number.
The function ${\mathcal R}(E_\mu,E_{th})$ is the average distance that a muon of
energy $E_\mu$ will travel until its energy reaches the value $E_{th}$, the
amount of energy needed to traverse the detector; this quantity is
expressed in the natural units for range, distance times the Earth's density.
$A_T(E_{th},\cos\theta_\nu)$ is the area projected onto a plane perpendicular to
the muon direction such that a muon of energy $E_{th}$ or greater can pass
through this part of the detector. The details for calculating
$A_T(E_{th},\cos\theta_\nu)$ can be found in Ref.~\cite{Lipari:1998}. For the
stopping muons,
$A_S(E_{th},\cos\theta_\nu)=A(\ell_{min},\cos\theta_\nu)-A_T(E_{th},
\cos\theta_\nu)$, where $A(\ell_{min},\cos\theta_\nu)$ is the projected area of
the detector with a path length greater than $\ell_{min}$ taken to be 7 m by the
experimentalists. Note that there is only muon data as electrons/positrons
produced in the rock are unable to travel to the detector. The data covers the
angular region from $\cos\theta_\nu = 0$ to $\cos\theta_\nu = -1$, directions
where muon production from the rock exceeds the cosmic ray background. Since the
neutrinos can originate as either electron or muon neutrinos, we sum over the
two neutrino flavors $\alpha$ as in Eq.~(\ref{upmu}). The upgoing muon data is
binned in ten angular bins and not binned in energy, resulting in a total of 20
data points.

Analysis of these events is not as computationally intensive as the calculation
of the contained events because the forward scattering allows the integration
over the scattering angles and the muon energy $E_\mu$ to be
performed outside the fitting program.  The parameters being fit are contained
in $\tilde{\mathcal P}_{\alpha\mu}(E_\nu,\cos\theta_\nu )$ which is independent
of the muon energy and scattering angle. We have found it efficient to change
the integration over these variables to the Feynman scaling variables $x$ and
$y$, as is natural for the deeply inelastic region.

\subsection{Super-K Statistical Analysis}

\begin{figure}[ht]
\includegraphics*[width=3.3in]{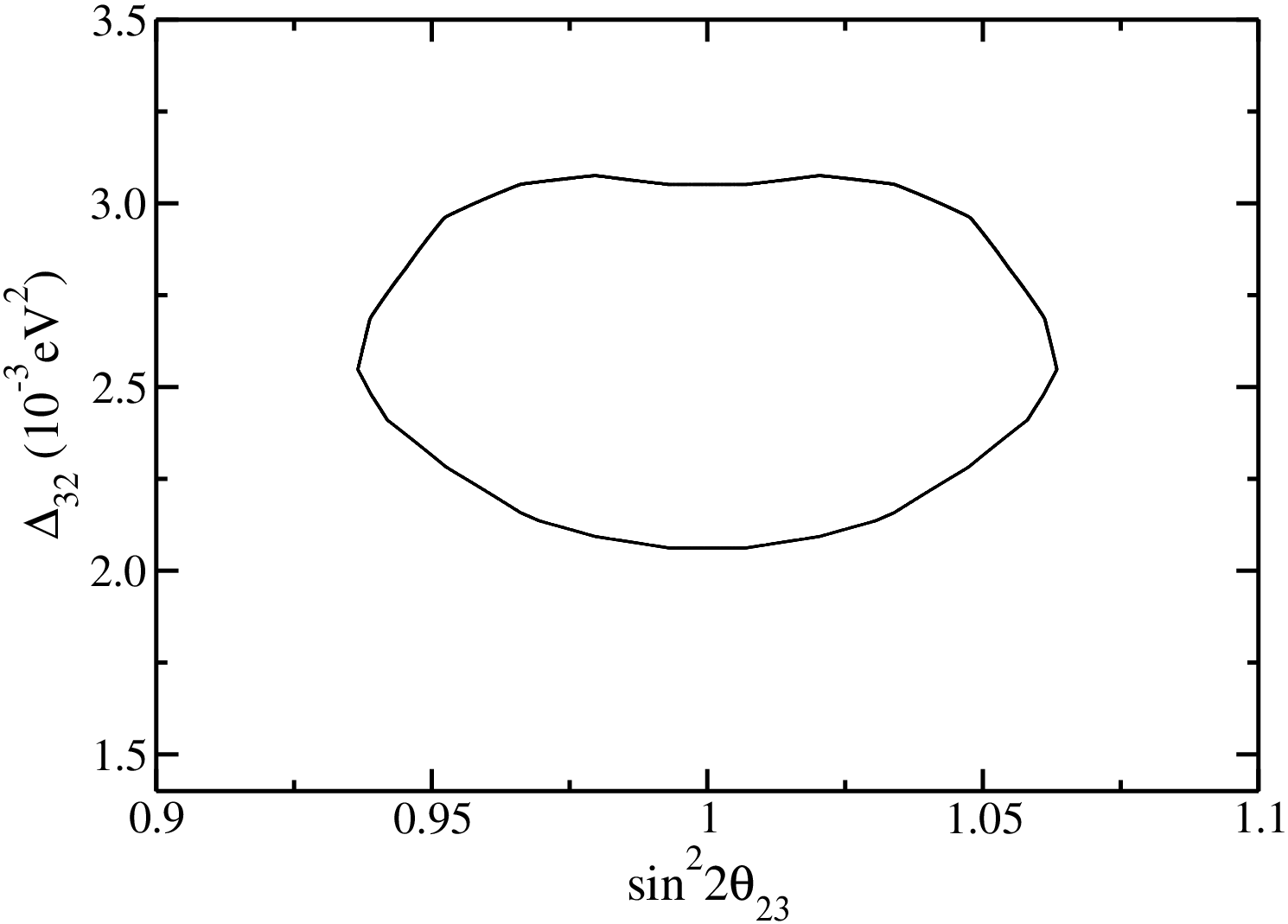}
\caption{Allowed region at 90\% confidence level for $\Delta_{32}$ and 
$\sin^{2}2\theta_{23}$ in the subdominant approximation using only Super-K
atmospheric data.}
\label{fig7}
\end{figure}

We use the most recent experimental data from Ref.~\cite{skatm4} which includes
180 data points for fully contained single-ring events, 90 for fully contained
multi-ring events, 80 for partially-contained events, 10 for upward
through-going muons, and  10  for upward stopping muons; in all, this
constitutes 370 data  points. In order to determine the neutrino  oscillation
parameters, we  construct a $\chi^2$ based upon a Poisson distribution,
following the  same  procedure used in Ref.~\cite{skatm4}.  We incorporate
systematic errors by utilizing the ``pull" approach as described in
Ref.~\cite{pulls}, which   allows one to incorporate  systematic errors in the
analysis without adding adjustable parameters. The approach is based upon
allowing linear corrections to the theoretical predictions for each systematic
error. Our $\chi^2$ function is
 \begin{eqnarray}
 \chi^2 &=& \sum_{n=1}^{370} \Bigg[ 2 \{ \overline{N}_{the}(n) - N_{obs}(n) \} 
 \nonumber \\ && + 2 N_{obs}(n) \ln \left(
\frac{N_{obs}(n)}{\overline{N}_{the}(n) } \right)  \Bigg]  
 + \sum_{i=1}^{43}  \left( \frac{\xi_i}{\sigma_i} \right)^2. \nonumber \\
 \label{chisq}
 \end{eqnarray}
$N_{obs}(n)$ is the number of observed events in the bin $n$; $N_{the}(n)  $ is
the theoretical prediction of the number of events in that bin; $\xi_i$ is the
systematic error pull for the systematic error $i$; and $\sigma_i$ is the
one-sigma value for the systematic error $i$. $\overline{N}_{the}(n)$ represents
a modified prediction of the expected number of events due to the inclusion of
systematic errors; the systematic errors adjust this quantity through an assumed
linear dependence on the pulls $\xi_i$.  Here we use 45 systematic errors
arising from different inputs into the data analysis as described in Tables
VII--X  taken from Ref.~\cite{skatm3}. For these 45 errors, all of them
contributed to the $\chi^2$ except the overall flux normalization and the
normalization for the multi-GeV multi-ring sample,  which are floated freely.
During each  fit, these forty-five  $\xi_i$  are   varied to minimize $\chi^2$
for a given  set of oscillation parameters. The  minimization of $\chi^2$ with
respect to $\xi_i$ ($\frac{\partial\chi^2}{\partial\xi_{j}}=0$) is equivalent in
the pull method to numerically solving for $\xi_i$ in the 45 coupled equations
of the form
 \begin{eqnarray}
\frac{\partial\chi^2}{\partial\xi_{j}}&=& \sum_{n=1}^{370}f_j^n\left( N_{th}(n)-
\dfrac{N_{obs}(n)}{1+\sum_{i=1}^{45} f_i^n \cdot   \xi_i} \right)\nonumber\\ 
 &&+ \sum_{i=1}^{43}  \frac{\xi_i}{\sigma^2_i} \delta_{ij}=0\,,
 \label{solvechisq}
 \end{eqnarray}
where $f_i^n $ is the $i$th systematic error for bin $n$. These equations are
linearized and solved as a set of coupled linear equations.

\begin{figure}[ht]
\includegraphics*[width=3.3in]{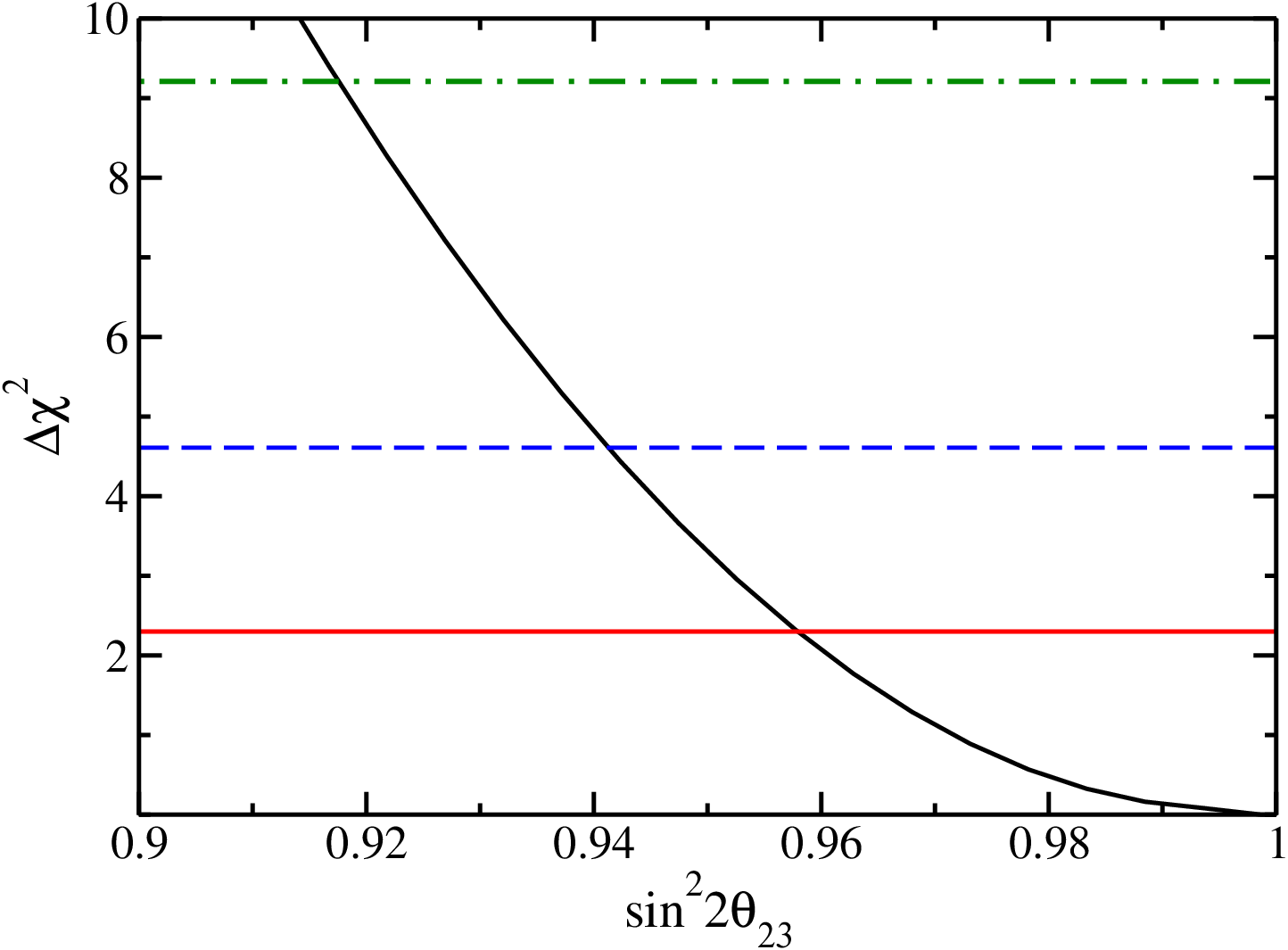}
\caption{[color online] $\Delta \chi^2$ versus $\sin^2 2\theta_{23}$ for our
analysis 
of the recent Super-K atmospheric data in the sub-dominant approximation. The 
horizontal lines take the values: [red] solid, $\Delta \chi^2=2.30$; [blue]
dashed, 
$\Delta \chi^2=4.61$; and [green] dot-dashed, $\Delta \chi^2=9.21$, the 68\%, 
90\%, and 99\% confidence levels for a two parameter fit.  Both $\Delta_{32}$ 
and $\theta_{13}$ are varied in obtaining this curve.}
\label{fig8}
\end{figure} 

\begin{figure}[ht]
\includegraphics*[width=3.3in]{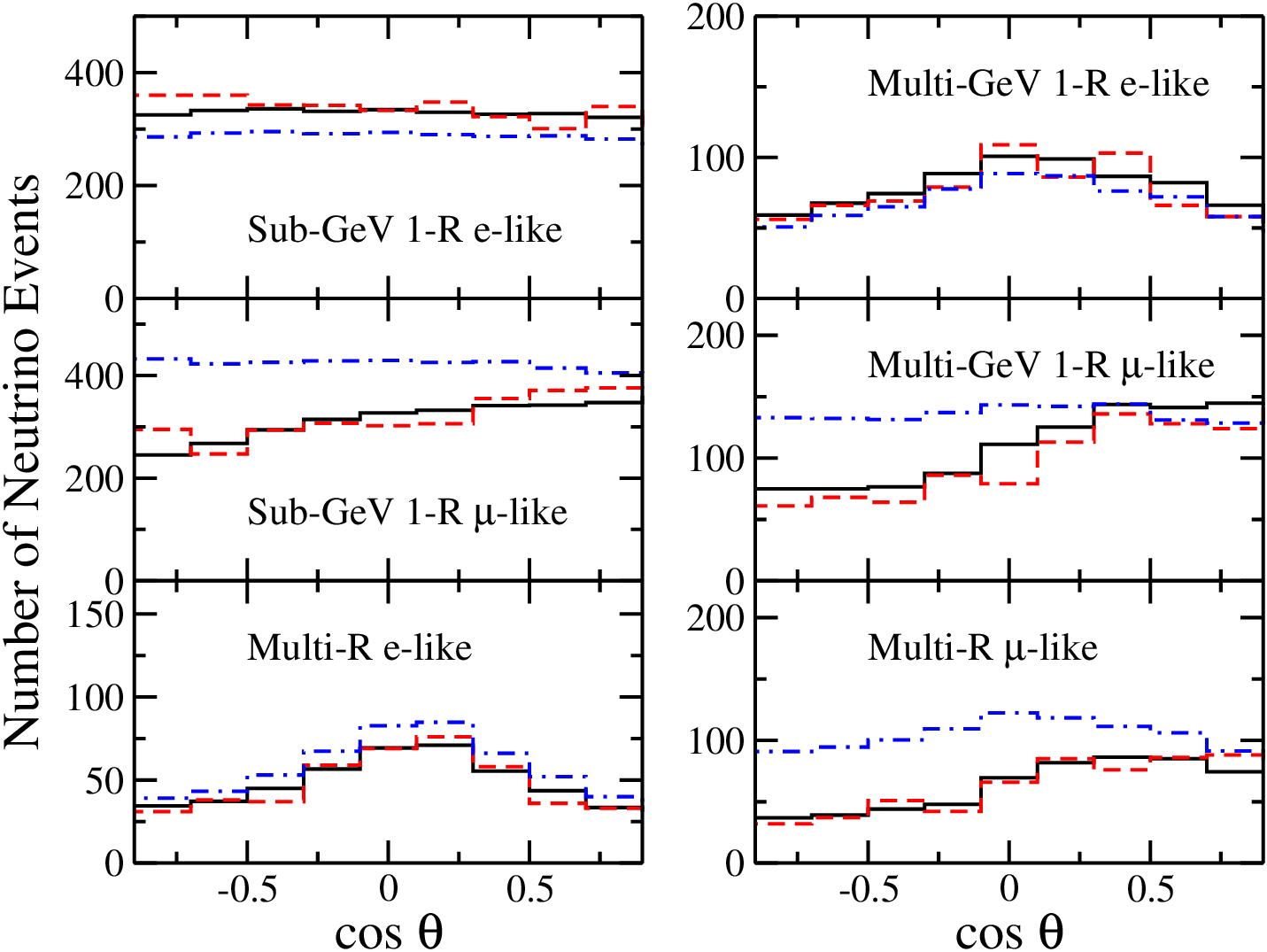}
\caption{The top two graphs depict the number of events for the sub-GeV data as
a function of the angular bin in $\cos\vartheta$, the middle two graphs are the
same except for the multi-GeV data, and the bottom two graphs are for the
multi-ring events.  The [black] solid curves represent the results from our best
fit parameters, the [red] dashed curves represent the data, and the [blue]
dot-dashed curves represent the Monte Carlo no-oscillation predictions. }
\label{fig9}
\end{figure}

We compare our analysis with that performed by the experimentalists in
Ref.~\cite{skatm4}. To do this, we utilize the sub-dominant approximation in
Eq.~(\ref{subd}) and minimize the above $\chi^2$. Our best fit oscillation
parameters  are
$(\Delta_{32},\sin^{2}\theta_{23},\theta_{13})=(2.5\times10^{-3}\text{
eV}^2,0.51,0.01)$ with an overall $\chi^2$ of 416 for the 370 data points. In
Fig.~1.a, the [black] solid curve represents $\Delta\chi^2$ versus $\Delta_{32}$
for the subdominant approximation, using only atmospheric data. In
Fig.~\ref{fig7}, we present the allowed region for $\Delta_{32}$ and 
$\sin^{2}2\theta_{23} $ corresponding to  $\Delta\chi^{2}=4.61 $, the 90\%
confidence level for a two parameter fit. 
We also present $\Delta\chi^2$ versus $\sin^{2}\theta_{23}$ in Fig.~\ref{fig8}.
 
At the 90\% confidence level for a two parameter fit, we find the allowed
parameter values $2.1\times10^{-3}\text{ eV}^2< \Delta_{32} < 3.1\times10^{-3}
\text{ eV}^{2}$  and $0.938<\sin^{2}2\theta_{23}$. Additionally, we extract the
allowed value for $\theta_{13}$ from the [black] solid curve in Fig.~3a,
$-0.38<\theta_{13}<0.38$. This is exactly the result of Ref.~\cite{skatm4},
and our other results are in excellent agreement with that analysis. 
 As noted previously, our reproduction of the allowed region
for $\theta_{13}$, which has a nonzero but small effect on the atmospheric data,
is a very strong test of our analysis.

\begin{figure}[ht]
\includegraphics*[width=3.3in]{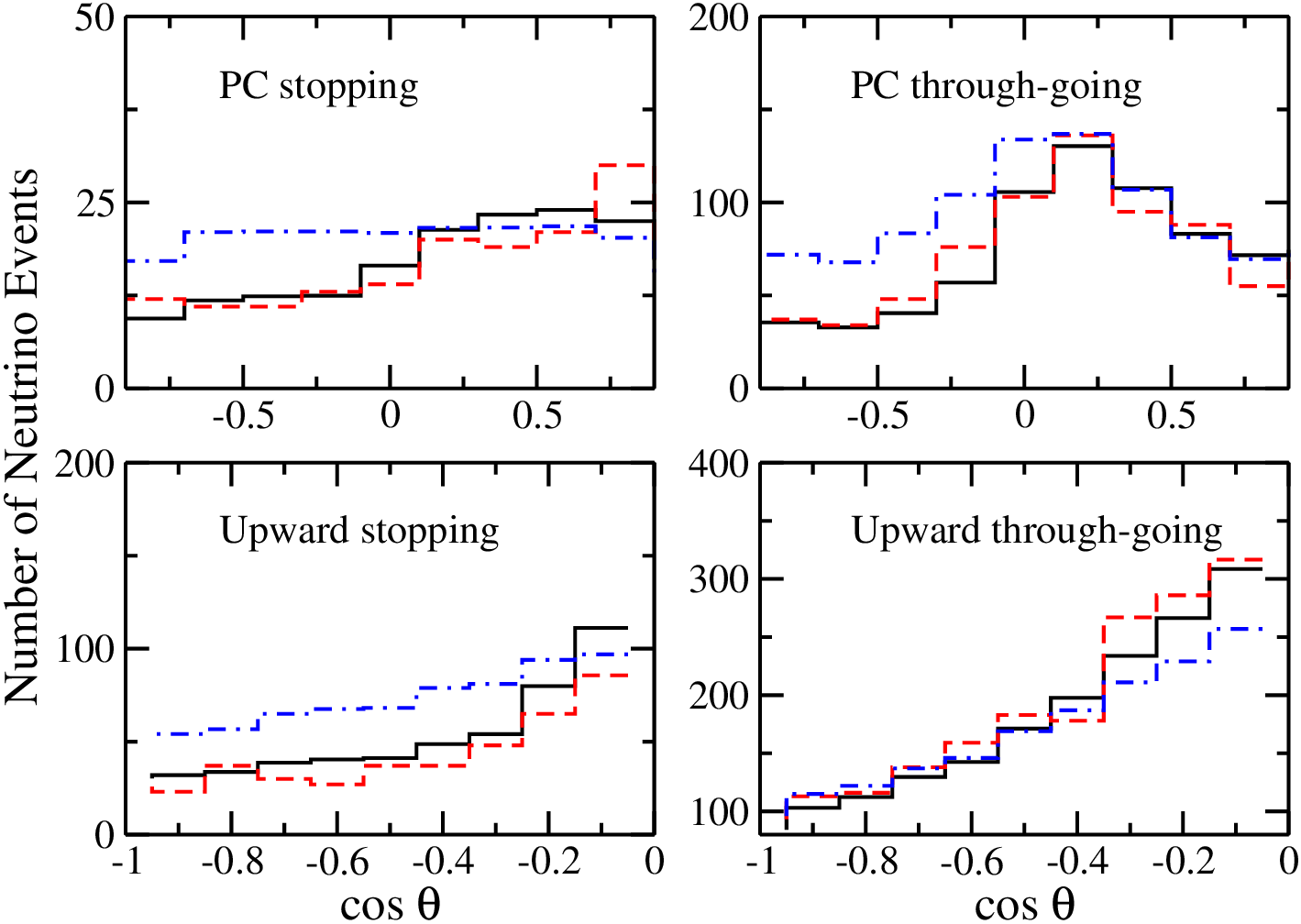}
\caption{The same as Fig.~\protect\ref{fig9} except the data sets are now the
partially contained stopping events (upper left), the partially contained
through-going events (upper right), the upward stopping muon events (lower
left), and the upward through-going muon events (lower right).}
\label{fig10}
\end{figure}

Finally, in Figs.~\ref{fig9} and \ref{fig10}, we compare the predicted number of
neutrino events corresponding to our best  fit parameters with the experimental
data as a function of the zenith angle.  We also present the Monte Carlo
predictions for the expected number of events in the absence of neutrino
oscillations. Each of the different Super-K atmospheric data sets is depicted.
The results are a good fit to the data, comparable to that found in
Ref.~\cite{skatm4}.

\subsection{CHOOZ experiment}

For the CHOOZ  reactor experiment,  we follow a standard procedure as  described
in Ref.~\cite{chooz}. In our analysis, we use experimental  data that consists
of seven positron energy bins for each of the two reactors, giving a total of 14
bins.  We include a  $14\times14$ covariance  matrix, $V_{ij}^{-1}$, to account
for the correlation between the energy bins, and we include the systematic error
from the overall normalization and energy calibration.   We write the  expected
positron yield for the $k^{th}$ reactor and the $j^{th}$ energy spectrum bin as
\begin{eqnarray}
  &&\overline{X}(E_j,L_k,\theta,\Delta_{32}) = \tilde{X}(E_j) 
  \overline{P}(E_j,L_k,\theta,\Delta_{32}),~~~\nonumber\\ 
&&~~~~~~~~~~(j=1,\ldots,7,~k=1,2)\,,
  \label{xosc}
\end{eqnarray}
where $\tilde{X}(E_j)$ is the distance-independent positron yield in the absence
of neutrino oscillations, $L_k$ is the reactor-detector distance, and 
$\overline{P}(E_j,L_k,\theta,\Delta_{32}) $ is the oscillation  probability
averaged over the energy bin and the  detector and reactor core sizes.
In our fitting routine, we  minimize the following $\chi^2$ function with
respect to the  neutrino oscillation parameters
\begin{widetext}
\begin{eqnarray}
&&\chi^2 (\theta,\Delta_{32},\alpha,g ) = \nonumber\\
&&~~~\sum_{i=1}^{14} \sum_{j=1}^{14}
\Bigl( X_i - \alpha \overline{X} \bigl( g E_i,L_i,\theta,\Delta_{32}
\bigr)\Bigr)  V_{ij}^{-1}
\Bigl( X_j - \alpha \overline{X} \bigl( g E_j,L_j,\theta,\Delta_{32}
\bigr)\Bigr) +
\left(\frac{\alpha-1}{\sigma_\alpha}\right)^2+\left(\frac{g-1}{\sigma_g}
\right)^2,
\label{chiA}
\end{eqnarray}
\end{widetext}
with the absolute normalization constant $\alpha$ and the energy-scale
calibration factor $g$.

\subsection{K2K experiment}

\begin{figure}
\includegraphics*[width=3.3in]{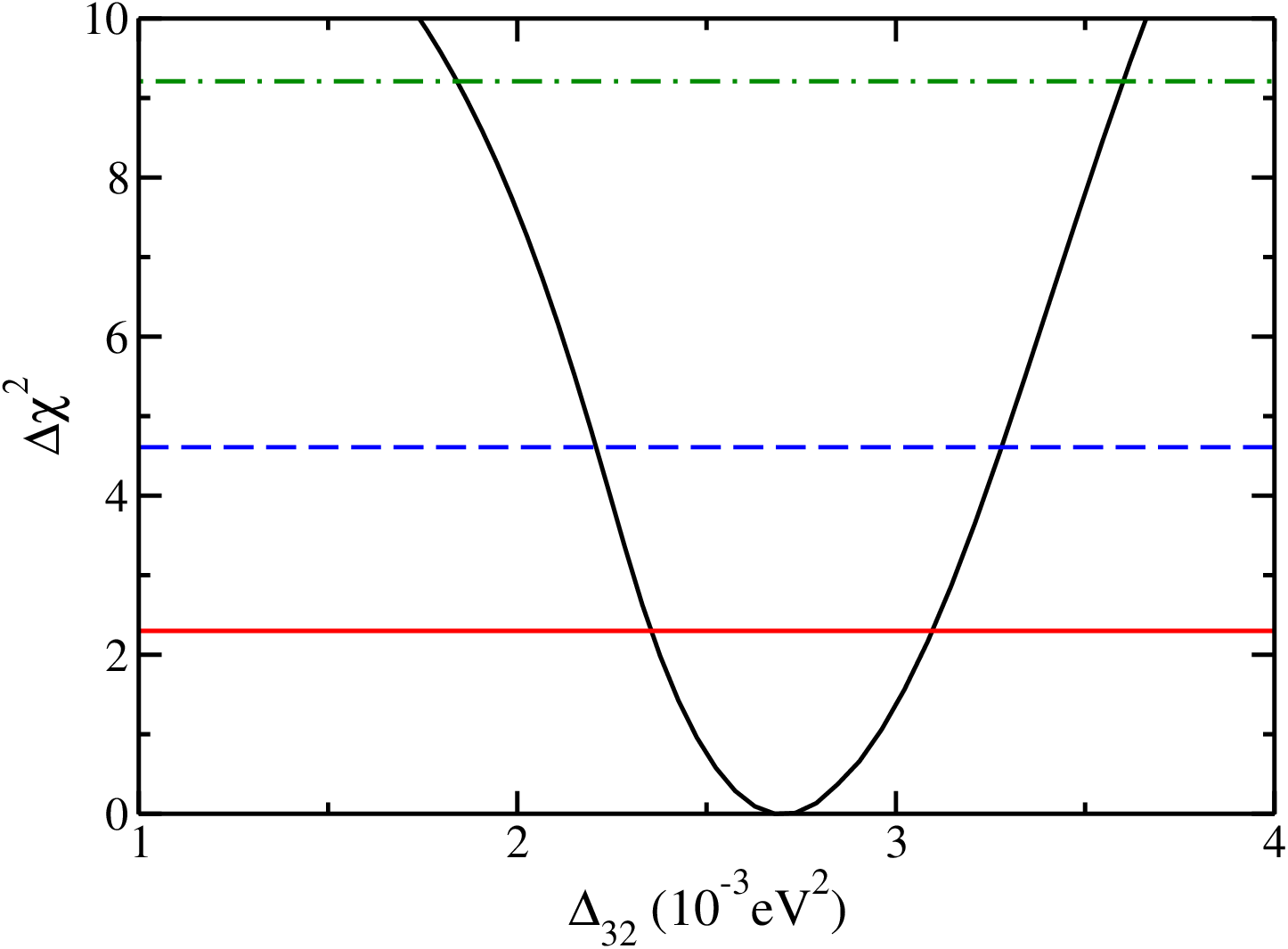}
\caption{[color online] $\Delta \chi^2$ versus $\Delta_{32}$ for our analysis 
of the K2K experiment in the sub-dominant approximation. The 
horizontal lines take the values: [red] solid, $\Delta \chi^2=2.30$; [blue]
dashed, 
$\Delta \chi^2=4.61$; and [green] dot-dashed, $\Delta \chi^2=9.21$, the 68\%, 
90\%, and 99\% confidence levels for a two parameter fit. }
\label{fig11}
\end{figure}

For the K2K experiment \cite{k2k06},  we employ the method developed in
Ref.~\cite{fogli:2002} to estimate  the  expected no-oscillation neutrino
spectrum, $S(E_{\nu}) $, in the relevant energy range of $\sim0.2$ to $\sim 3.0$
GeV. The expected number of neutrino events for oscillating neutrinos is then
\begin{eqnarray}
 N^{theo}_n= \int^{E_{max}(n)}_{E_{min}(n)} S(E_{\nu}) P_{\mu\mu}(L/E_{\nu})\,,
\label{k2kth}
\end{eqnarray}
\noindent
where $ P_{\mu\mu}(L/E_{\nu})$ is the muon neutrino survival probability and
$E_{max}(n)$ ($E_{min}(n)$) are the maximum (minimum) energy values for the
energy bin $n$.
For the statistical analysis, we follow the procedure described  in
Ref.~\cite{Masayathesis}.  We only use the single-ring  sub-sample, which
consist of 58 neutrino events.
The signature for  neutrino oscillations from $\nu_\mu$ to $\nu_{\tau}$ in a two
neutrino analysis are both a reduction in the total number of observed neutrino
events and a distortion in the neutrino
energy spectrum.  The $\chi^2$ function is divided into two terms:  the observed
total number of events detected at the Super-K detector, $\chi^{2}_{norm}$,
and the shape of the spectrum included in  $\chi^{2}_{shape}$. We use the
``pull'' method  \cite{pulls}  to account for 31 systematic uncertainties by
adding a third term $\chi^{2}_{syst}$.
\begin{eqnarray}
\chi^{2}_\mathrm{K2K} = \chi^{2}_\mathrm{norm}+
\chi^{2}_\mathrm{shape}+\chi^{2}_\mathrm{syst}\,.
\end{eqnarray}
The best fit oscillation parameters, $\Delta_{32}$ and $\theta_{23}$, are
obtained by minimizing  $\chi^{2}_\mathrm{K2K}$. 

The systematic parameters included in $\chi^{2}_{syst}$ arise from the neutrino
energy spectrum at the near detector site, the  flux ratio, the neutrino-nucleus
cross-section, the efficiency and the energy scale of the Super-K detector, and
the overall normalization.  The $k$th systematic error is represented by the
coefficient $C_n^k$ and modifies the expected number of neutrino events,
Eq.~(\ref{k2kth}), in a linear manner according to the ``pull" method
\begin{eqnarray}
\tilde N_n^\mathrm{theo}&=&N^\mathrm{theo}(n) + \sum_{k=1}^{31}C_n^k
\xi_k\,,\nonumber\\
\tilde N^\mathrm{theo}_\mathrm{total}&=&\sum_{n=1}^8 \tilde N_n^\mathrm{theo}\,,
\end{eqnarray}
with  $\xi_k$  the pull corresponding to systematic error $k$.

Due to the low statistics, we employ a Poisson distribution; hence, the
expressions  for $\chi^{2}_{norm}$ and $\chi^{2}_{shape}$
are given by
\begin{eqnarray}
\chi^2_\mathrm{norm}&=& 2\left( \tilde
N^\mathrm{theo}_\mathrm{total}-N^\mathrm{data}_\mathrm{total} -
N^\mathrm{data}_\mathrm{total}\ln
\frac{\tilde
N^\mathrm{theo}_\mathrm{total}}{N^\mathrm{data}_\mathrm{total}}\right)\,\,,
\nonumber\\
\chi^2_\mathrm{shape}&=& 2\sum_{n=1}^8
\left(\tilde N_n^\mathrm{theo}-N_n^\mathrm{data} - N_n^\mathrm{data}\ln
\frac{\tilde N_n^\mathrm{theo}}{N_n^\mathrm{data}}\right)\,,\nonumber\\
\label{chik2k}
\end{eqnarray}
where $N_n^\mathrm{data}$ is the  experimental data provided  by the K2K
collaboration \cite{k2k06} and the superscript ``total" implies a sum over $n$.
The contribution to $\chi^2$ from the systematic errors is
\begin{equation}
\chi^2_\mathrm{syst}=\sum_{j,k=1}^{31}\xi_k\,M_{kj}^{-1}\,\xi_j\,,
\end{equation}
where we use an error matrix $M_{kj} $  constructed from Tables 8.1 and 8.2
provided in Ref.~\cite{Masayathesis}.

\begin{figure}
\includegraphics*[width=3.3in]{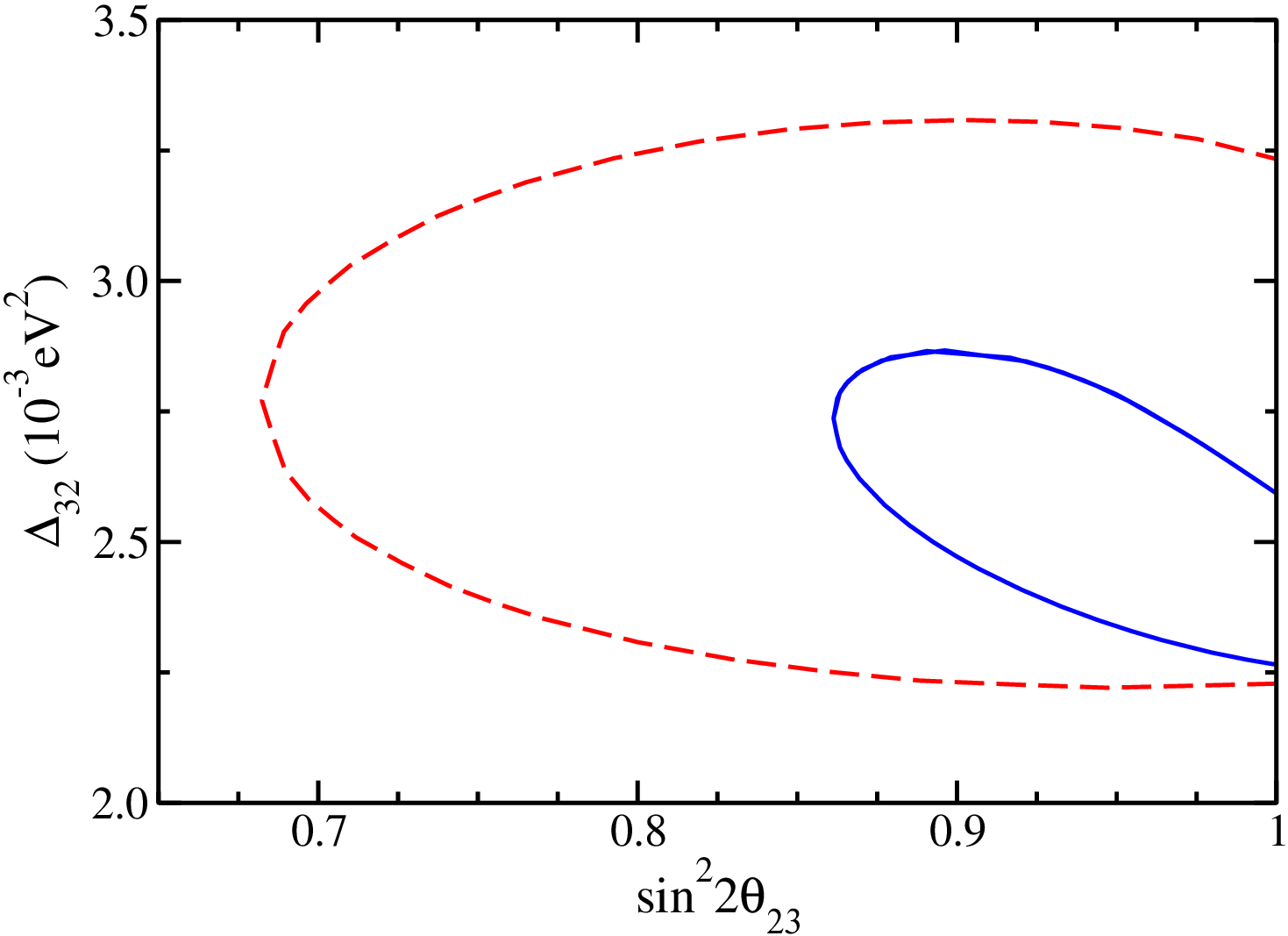}
\caption{[color online] $\Delta_{32}$  vs $\sin^{2}2\theta_{23}$  allowed region
for a 
two parameter fit in the subdominant approximation 
at the 90\% CL for the K2K data, [red] dashed, and the 
MINOS, [blue] solid, experiments.}
 \label{fig12}
\end{figure}

In Fig.~\ref{fig11}, we depict $\Delta \chi^2$ versus $\Delta_{32}$ for an
analysis that utilizes only the K2K data in the subdominant approximation;
also, Fig.~\ref{fig12} shows the 90\% CL allowed region in the [red] dashed
contour for $\Delta_{32}$ and $\sin^{2}2\theta_{23}$ in the subdominant
approximation. 
The absolute minimum in our fit is  $(\Delta_{32},
\sin^{2}2\theta_{23})=(2.78\times10^{-3}\text{ eV}^2,0.998)$.  At 90\% CL, we
find  $2.2\times 10^{-3} \text{ eV}^2 <\Delta_{32}  <3.2\times 10^{-3} \text{
eV}^2$.
The total number of observed events 58  is in agreement with the 56 events found
from the model. All of these results are  consistent with the analysis performed
by the experimentalists in Ref.~\cite{k2k06}.

\subsection{MINOS experiment}

\begin{figure}
\includegraphics*[width=3.3in]{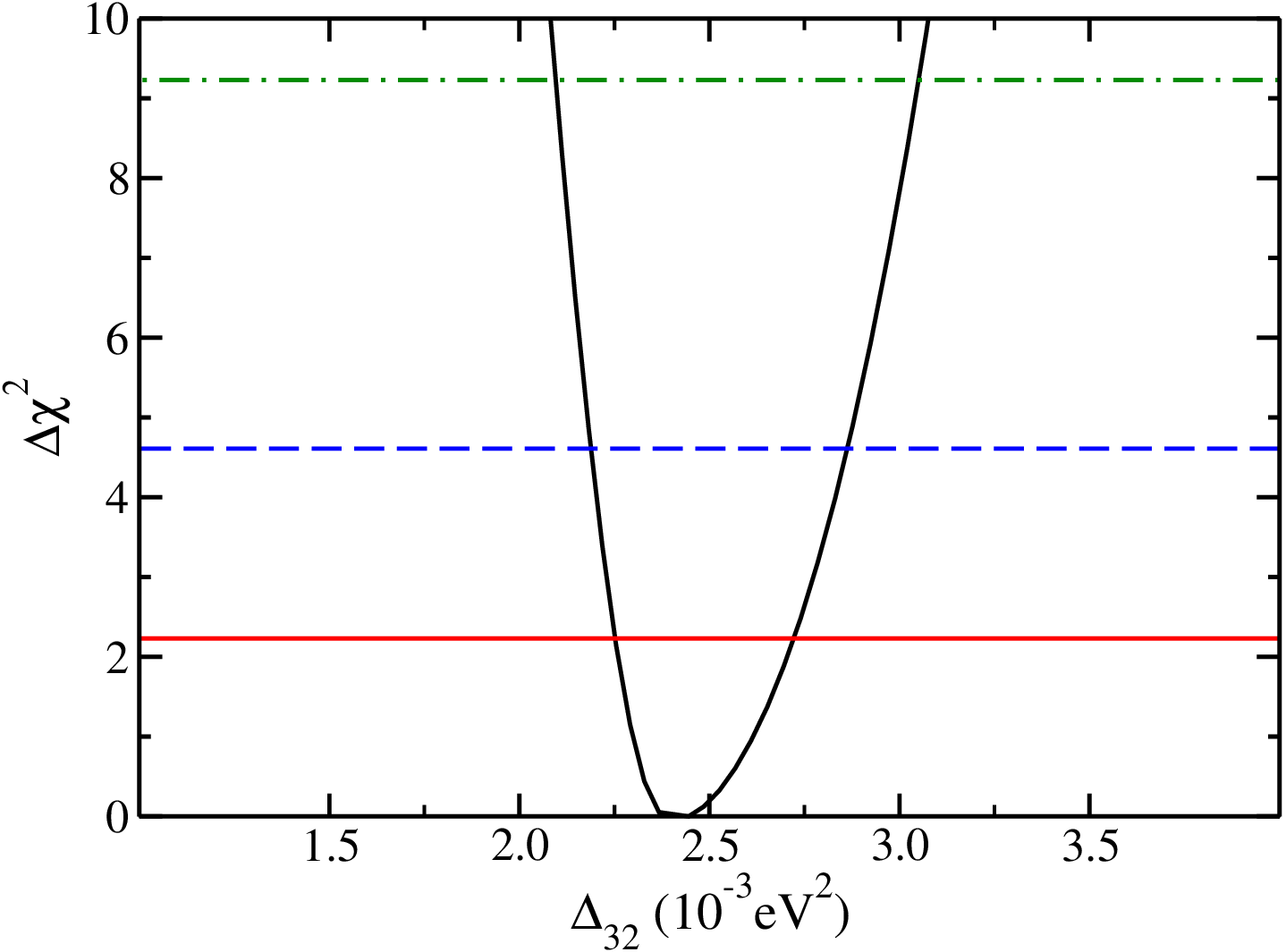}
\caption{[color online] The same as Fig.~\protect\ref{fig11} except the data are
from MINOS.}  \label{fig13}
\end{figure}

The MINOS experiment is quite similar to the K2K experiment and thus we apply a
similar analysis technique. For the no-oscillation spectrum, we use the Monte
Carlo simulation provided by the MINOS collaboration \cite{Adamson:2008}. We
normalized this spectrum to measurements made at the near detector. A total of
1065 events were expected in the absence of neutrino oscillations. With the
no-oscillation spectrum and Eq.~(\ref{k2kth}), we calculate the expected number
of neutrino events in the presence of neutrino oscillations. We use the MINOS
data \cite{Adamson:2008} corresponding to two years of  beam operation in which
884 $\nu_{\mu}$  neutrino events are observed.
The data consists of 15 energy bins along with three systematic errors:  the
relative normalization between the far and near detectors with a 4\%
uncertainty; the absolute hadronic energy scale with a 11\% uncertainty; and a
50\% uncertainty in the neutral-current background rate.
For the definition of $\chi^2$,  we use the Poisson distribution function
\begin{eqnarray}
\chi^2_\mathrm{MINOS}&=& 2\sum_{n=1}^{15}\left(\tilde
N_n^\mathrm{theo}-N_n^\mathrm{data} - N_n^\mathrm{data}\ln
\frac{\tilde N_n^\mathrm{theo}}{N_n^\mathrm{data}}\right) \nonumber \\
&& +\sum_{j=1}^{3}\left(\frac{\xi_{j}}{\sigma_{j}}\right)^{2},
\label{chiminos}
\end{eqnarray}
where the symbols are  analogously defined to those in the K2K section.

In Fig.~\ref{fig13} we plot $\Delta \chi^2$  versus $\Delta_{32}$ for the K2K
data in the subdominant approximation; likewise, in Fig.~\ref{fig12}, we present
the allowed region at 90\% CL, the [blue] solid curve, 
for  $\Delta_{32}$ and $\sin^{2}2\theta_{23}$. The
minimum  of $\chi^2$ is located at $\Delta_{32} =2.41\times 10^{-3}$ eV$^2$  and
$\sin^{2}2\theta_{23}=0.9990$. The allowed intervals of these parameters at 90
$\%$ CL  are $2.25\times 10^{-3}\text{ eV}^2<\Delta_{32} <2.8\times 10^{-3}
\text{ eV}^2$  and  $0.86<\sin^{2} 2\theta_{23}$ for $\Delta \chi^2=4.6$. 
All of these results are consistent with the analysis performed by the
experimentalists in Ref.~\cite{Adamson:2008}.

{Results which combine CHOOZ, K2K, and MINOS are presented in the main body of
this work. This Appendix provides details of the analysis tools we use
throughout this work; additional details may be found in Ref.~\cite{jesus}.

\section{ACKNOWLEDGMENTS}

The work of J.~E.~R.~and D.~J.~E.~is supported, in part, by US Department of
Energy Grant DE-FG02-96ER40975. The work of J.~E.~R. is also supported, in part, 
by CONACyT, Mexico. The work of D.~C.~L.~is supported, in part, by
US Department of Energy Grant DE-FG02-96ER40989.

\bibliography{biblio}

\end{document}